\documentclass[journal]{IEEEtran}

\usepackage{fancyhdr}
\usepackage{amsmath,amssymb}
\usepackage{amsfonts}
\usepackage[dvips]{graphicx}
\usepackage{dsfont}
\usepackage{balance}
\usepackage{algpseudocode}
\usepackage{algorithm}

\usepackage[hidelinks]{hyperref}    

\usepackage{color}
\usepackage{pgfplots}           
\pgfplotsset{compat=1.16}

\usepackage{enumerate}
\usepackage{graphics}
\usepackage{subfigure}
\usepackage{cite}
\usepackage{mathrsfs}
\usepackage{stfloats}

\usepackage{tikz}
\usepackage{mathdots}
\usepackage{yhmath}
\usepackage{cancel}
\usepackage{color}
\usepackage{siunitx}
\usepackage{array}
\usepackage{multirow}
\usepackage{textcomp}
\usepackage{gensymb}
\usepackage{tabularx}
\usepackage{booktabs}
\usetikzlibrary{fadings}
\usetikzlibrary{patterns}
\usetikzlibrary{shadows.blur}

\newtheorem{theorem}{Theorem}

\newtheorem{corollary}{Corollary}

\newtheorem{definition}{Definition}

\newtheorem{proposition}{Proposition}
\newtheorem{remark}{Remark}

\begin{document}

	\title{New Results for Pearson Type III \\ Family of Distributions and Application \\ in Wireless Power Transfer}
	\author{
		Sotiris A. Tegos,~\IEEEmembership{Student Member,~IEEE,}
		George K. Karagiannidis,~\IEEEmembership{Fellow,~IEEE,}
		Panagiotis D. Diamantoulakis,~\IEEEmembership{Senior Member,~IEEE,} 
		and	Nestor D. Chatzidiamantis,~\IEEEmembership{Senior Member,~IEEE}
		\thanks{S. A. Tegos, G. K. Karagiannidis, P. D. Diamantoulakis, and N. D. Chatzidiamantis are with Wireless Communication and Information Processing  Group  (WCIP), Department of Electrical and Computer Engineering, Aristotle University
of Thessaloniki, 54 124, Thessaloniki, Greece (e-mails: \{tegosoti,geokarag,padiaman,nestoras\}@auth.gr).}
	}
	\maketitle

\begin{abstract}
Pearson and log Pearson type III distributions have been considered in several scientific fields, as in hydrology and seismology. In this paper, we present new results for these distributions and we utilize them, for first time in the literature, to investigate the statistical behavior of  wireless power transfer (WPT), assuming that the harvested energy follows a well-established nonlinear energy harvesting model based on the logistic function. 
Specifically, we present new closed-form expressions for the statistical properties of a general form of Pearson and log Pearson type III distributions and we utilize them to introduce a new member of the Pearson type III family, the logit Pearson type III distribution, through which the logit gamma distribution is also defined. 
Moreover, we derive closed-form expressions for the probability density function, the cumulative distribution function and moments of the distributions of the sum, the log sum and the logit sum of Pearson type III  random variables. 
Furthermore, taking into account that Pearson type III family of distributions is closely related to the considered nonlinear harvesting model the statistical properties of the distribution of the harvested power and derived, for both single input single output and multiple input single output scenarios.
\end{abstract}


\section{Introduction}

The Pearson type III and log Pearson type III distributions \cite{Johnson1995,Singh1998_p,Singh1998_lp} attracted the interest of the research community, since they have been utilized in several scientific fields, such as hydrology and seismology. Specifically, they are frequently used in hydrology for flood frequency analysis \cite{Bobee1975}, while in \cite{Phien1984} the log Pearson type III distribution was applied to flood and maximum rainfall data and its general use in fitting annual rainfall and streamflow sequences, was investigated. 
Furthermore, in \cite{Gupta1994} it was found that the log Pearson type III distribution can effectively describe  the behavior of the maximum earthquake magnitudes for all ranges and also be applied to evaluate the design magnitudes.
Although the Pearson type III and log Pearson type III distributions have been investigated in the existing literature, a more general form of these distributions is not fully investigated. Moreover, the distributions of the sum and the log sum of Pearson type III random variables (RVs) have not been examined.

Regarding communication systems, Pearson type III distribution can be considered as a generalized form of the gamma distribution, which is frequently used in wireless communications when the channel fading is assumed to follow  Nakagami-\textit{m} distribution. 
Also, in \cite{Shi2017} the outage performance of hybrid automatic repeat requests with incremental redundancy (HARQ-IR) was investigated, through the cumulative distribution function (CDF) of the product of multiple correlated shifted gamma RVs, which is a special case of the Pearson type III distribution. Finally, in \cite{Kim2003} the shifted gamma distribution is used to model long-range dependent internet traffic, when the input traffic rate is not Gaussian.

In this paper, we utilize the Pearson type III family distribution in wireless power transfer (WPT) and especially in investigating the statistical properties of the harvested energy, which is a prerequisite in order to analytically evaluate  the capabilities and reliability of this technology. It should be highlighted that energy harvesting (EH) is a promising solution for prolonging the lifetime of Internet-of-Things (IoT) networks by offering self-sustainability to the devices, minimizing -if not eliminating- the use of battery power. This is of paramount importance  especially when replacing or recharging the batteries is inconvenient, costly, or dangerous, such as in remote areas, harsh industrial environments, e.g., rotating and moving platforms, human bodies, or vacuum equipment \cite{diamantoulakis2018resource}. However, the main disadvantage of basic EH methods is their reliability, since they depend solely on ambient natural energy sources, such as wind and solar, which are uncontrollable. To this end, WPT which utilizes radio frequency (RF) signals for EH is an interesting alternative and also benefits from high-density networks \cite{Grover2010,Krikidis2012,Krikidis2014,Krikidis2020}.

Scanning the open literature, a linear EH model was used to express the harvested energy, when WPT is performed \cite{Tegos2020, Zhou2014, Diamantoulakis2016, Zhang2013}. This model can be easily handled because of its simplicity, however, it can be considered impractical, since it is not accurate and cannot describe the saturation of the EH circuit.  Nonlinear EH models were proposed in \cite{Chen2017,Clerckx2018}. Also, in \cite{Boshkovska2015} a practical parametric nonlinear EH model was proposed and its accuracy was verified through measurements. This model is based on the logistic function and due to its accuracy has been adopted in several research works \cite{Kang2018, Clerckx2019, Tegos2019, Jiang2020}.
However, although this nonlinear EH model has received the researchers' attention, a comprehensive analytical framework which can assist in the design and evaluation of its performance has not been yet provided and its statistical properties, e.g., the CDF, the probability density function (PDF) and the moments, have not been derived. Considering that the basis of this model is the logistic function, the analytical investigation of WPT performance is facilitated by the use of logit Pearson type III distribution, which however has not been defined and studied in existing literature, in which solely the  the logit normal distribution has been considered \cite{Johnson1949,Frederic2008}.


\subsection{Contribution}

In the present work, we introduce the logit Pearson type III distribution and we utilize the statistical properties of this distribution to investigate the performance of WPT systems where the nonlinear EH model proposed in \cite{Boshkovska2015} is considered.
The specific contributions of this paper are listed below:
\begin{itemize}
	\item We introduce a new member of the Pearson type III family, the logit Pearson type III distribution, and derive closed-form expressions for its statistical properties, e.g., the CDF, the PDF and the moments. To this end, firstly, we provide new results for a general form of the Pearson type III and log Pearson type III distributions.
	\item We derive closed-form expressions for the statistical properties of the distribution of the sum of Pearson type III,  log   Pearson type III and  logit  Pearson type III RVs.
	\item We utilize the above new statistical results to provide a comprehensive analytical framework for the evaluation of the performance of the EH systems, when the considered nonlinear EH model is used, and to analytically evaluate the capabilities and reliability of WPT technology. Useful insights for the EH system can be extracted through the evaluation of  the average harvested power and harvested power probability of outage.  Both  single input single output (SISO) and multiple input single output (MISO) scenarios are considered. Specifically, for the MISO scenario two cases are considered, i.e., a network with a power beacon (PB) with multiple antennas and a network with multiple PBs with a single antenna. 
\end{itemize}

\subsection{Structure}

The remainder of the paper is organized as follows: \\
In Section II, the statistical properties of the Pearson type III and the log Pearson type III distribution are derived and the distributions of the sum of Pearson type III and the log sum of Pearson type III RVs are investigated.
In Section III, the logit Pearson type III distribution is introduced and its statistical properties are derived as well as the ones of the sum of logit Pearson type III distributions.
In Section IV, the expressions for the CDF, the PDF and the moments of the harvested energy are derived considering the nonlinear EH model for the SISO and the MISO scenario.
Finally, closing remarks and discussions are provided in Section V.

\section{New Results for the Pearson and the Log Pearson Type III Distributions}

In this section, the Pearson type III and log Pearson Type III distributions are presented and new results, regarding their statistical properties, are provided. Also, the distribution of the sum of Pearson type III and  log Pearson type III RVs are investigated.

\subsection{The Pearson Type III Distribution}

If a RV $X$ follows the Pearson type III distribution with parameters $(a,b,m)$, where $a \in \mathbb{R}$ with $a > 0$ is the shape parameter, $b\in \mathbb{R}$ with $b\neq 0$ is the inverse scale parameter and $m\in \mathbb{R}$ is the shift parameter, then its PDF is given by \cite{Bobee1975}
\begin{equation} \label{PT3_PDF}
	f_X(x,a,b,m) = \frac{\left| b\right|}{\Gamma(a)} (b (x-m))^{a-1} e^{-b (x-m)},
\end{equation}
where $\Gamma(\cdot)$ is the gamma function and $e$ is the base of the natural logarithm. If $b>0$, $x\in (m,+\infty)$ and if $b<0$, $x\in (-\infty,m)$. 

In the following proposition, the CDF of the Pearson type III distribution is provided for $b<0$. For $b>0$, the CDF is provided in \cite{Singh1998_p}, but it is also included in the following proposition for completeness.
\begin{proposition} \label{prop_cdf_p}
	The CDF of the Pearson type III distribution can be expressed as
	\begin{equation} \label{PT3_CDF}
	F_X(x,a,b,m) = \begin{cases}
	\frac{1}{\Gamma(a)} \Gamma (a, b (x - m)), &   b < 0 \\
	\frac{1}{\Gamma(a)} \gamma (a, b (x - m)), &   b > 0 \text{ \cite{Singh1998_p}},
	\end{cases}
	\end{equation}
where $\Gamma(\cdot,\cdot)$ and $\gamma(\cdot,\cdot)$ is the upper and lower incomplete gamma function, respectively.
\end{proposition}

\begin{IEEEproof}
	If $b<0$, $x\in (-\infty,m)$ the CDF is given by
	\begin{equation} \label{CDF1}
	F_X(x,a,b,m) = -\frac{b}{\Gamma(a)} \int_{-\infty}^x (b (y-m))^{a-1} e^{-b (y-m)} dy.
	\end{equation}
	By using $z = b (y-m)$, \eqref{CDF1} can be written as
	\begin{equation}
	F_X(x,a,b,m) = \frac{b}{\Gamma(a)} \int_{b (x-m)}^{\infty} z^{a-1} e^{-z} dy.
	\end{equation}
	Considering the definition of the upper incomplete gamma function, the expression of the CDF when $b<0$ is derived and the proof is completed.
\end{IEEEproof}

In the next proposition, the moments of the Pearson type III distribution are provided. When $b>0$, the expression for the moments is presented in \cite{Singh1998_p}. We extract the same expression when $b<0$.
\begin{proposition}
	The $n$-th moment of the Pearson type III distribution is given by
	\begin{equation} \label{PT3_moments}
	\mu_X^n (a,b,m) = \sum _{k=0}^n \binom{n}{k} \frac{m^{n-k} \Gamma(k+a)}{b^k \Gamma(a)},
	\end{equation}
	where $\binom{n}{k}$ denotes the binomial coefficient.
\end{proposition}

\begin{IEEEproof}
	If $b<0$, the $n$-th moment is calculated as
	\begin{equation} \label{p_mom_pr}
	\mu_X^n (a,b,m) = \int_{-\infty}^m x^n f_X(x,a,b,m) dx.
	\end{equation}
	By changing variable and utilizing the binomial theorem, \eqref{p_mom_pr} can be rewritten as
	\begin{equation}
	\mu_X^n (a,b,m) = \frac{1}{\Gamma(a)} \sum_{k=0}^{n} \binom{n}{k} \frac{m^{n-k}}{b^k} \int_0^{\infty} z^{k+a-1} e^{-z} dz.
	\end{equation}
	From the definition of gamma function, \eqref{PT3_moments} is derived. 
\end{IEEEproof}


\begin{corollary}
	From \eqref{PT3_moments}, the mean value is obtained as the first moment and can be expressed as
	\begin{equation}
	\mu_x^1 = \frac{a}{b} + m.
	\end{equation}
\end{corollary}


The characteristic function of the Pearson type III distribution is given by \cite{Bobee1975}
\begin{equation}
\phi_X(t) = \frac{e^{j m t}}{\left(1-\frac{j t}{b}\right)^{a}},
\end{equation}
where $j^2=-1$.

\begin{remark}
	The gamma distribution is a special case of the Pearson type III distribution with $b>0$ and $m=0$. 
	If $b\neq 0$, a RV that follows the Pearson type III distribution with parameters $(a,b,m)$ can be also multiplied with a constant $c$ resulting in a RV that follows the Pearson type III distribution with parameters $(a,\frac{b}{c},m c)$. Accordingly, if a RV follows the gamma distribution with parameters $(a,b)$ with $b>0$, multiplying with a negative constant $c$ in a RV that follows the Pearson type III distribution with parameters $(a,\frac{b}{c},0)$, where the second parameter is negative.
\end{remark}

\subsection{The Log Pearson Type III Distribution}

If the RV $X$ follows the Pearson type III distribution with parameters $(a,b,m)$, the RV $Y = e^X$ follows the log Pearson type III distribution with the same parameters.
If $b>0$, $y\in (e^m,+\infty)$ and if $b<0$, $y\in (0,e^m)$. When $b<0$, the distribution can also be considered as inverse log Pearson type III.


The PDF of the log Pearson type III distribution is given by \cite{Bobee1975}
\begin{equation} \label{lP_PDF}
f_Y(y,a,b,m) = \frac{|b| e^{bm}}{\Gamma(a)} (b(\ln y - m))^{a-1} y^{-b-1},
\end{equation}
where $\ln(\cdot)$ is the natural logarithm.

In the following proposition, the CDF of the log Pearson type III distribution is provided for $b<0$. For $b>0$, the CDF is provided in \cite{Singh1998_lp}, but it is also included in the following proposition for completeness.
\begin{proposition}
	The CDF of the log Pearson type III distribution can be expressed as
	\begin{equation} \label{lP_CDF}
	F_Y(y,a,b,m) = \begin{cases}
	\frac{1}{\Gamma(a)} \Gamma (a, b (\ln y - m)), &   b < 0 \\
	\frac{1}{\Gamma(a)} \gamma (a, b (\ln y - m)), &   b > 0 \text{\cite{Singh1998_lp}}.
	\end{cases}
	\end{equation}
\end{proposition}

\begin{IEEEproof}
	The CDF of the log Pearson type III distribution is derived by integrating \eqref{lP_PDF} or directly from \eqref{PT3_CDF}.
\end{IEEEproof}


The $n$-th moment of the log Pearson type III distribution is given by \cite{Bobee1975}
\begin{equation} \label{LPT3_moments}
\mu_Y^n (a,b,m) = e^{mn} \left( \frac{b}{b-n} \right)^a.
\end{equation}
If $b>0$, $b>n$ should be satisfied.
If $b<0$, there is no constraint.



\begin{corollary}
	From \eqref{LPT3_moments}, the mean value is obtained as the first moment and can be expressed as
	\begin{equation}
	\mu_Y^1 = \frac{a}{b} + m.
	\end{equation}
\end{corollary}

An approximation of the characteristic function of the log Pearson type III distribution is provided in the following proposition.
\begin{proposition}
	The characteristic function of the log Pearson type III distribution is approximately given as a formal power series \cite{Karagiannidis2004} by
	\begin{equation} \label{LPT3_char}
	\phi_Y(t) = \sum _{n=0}^\infty \frac{\mu_n}{n!} (j t)^n = \sum _{n=0}^\infty \frac{e^{m n}}{n!} \left( \frac{b}{b-n} \right)^a (j t)^n,
	\end{equation}
	where $n!$ is the factorial of $n$. \eqref{LPT3_char} stands only when $b<0$.
\end{proposition}

\begin{corollary}
	The infinite series in \eqref{LPT3_char} converges.
\end{corollary}

\begin{IEEEproof}
By setting $a_n = \frac{e^{m n}}{n!} \left( \frac{b}{b-n} \right)^a \! (j t)^n$, it holds that 
	\begin{equation}
	\lim_{n \rightarrow \infty} \left| \frac{a_{n+1}}{a_{n}} \right| = \lim_{n \rightarrow \infty} \left| \frac{e^m j t}{n+1} \left( \frac{b-n}{b-n-1} \right)^a \right|
	= 0.
	\end{equation}
\end{IEEEproof}

\subsection{Distribution of the Sum of Pearson Type III RVs}

In this subsection, the distribution of the sum of Pearson type III RVs is investigated.
Let $\left\{X_q\right\}_{q=1}^{L}$ be a set of $L$ RVs following the Pearson type III distribution with $a_i \in \mathbb{Z}$, $a>0$ and either $b_i > 0$, $\forall i$ or $b_i < 0$, $\forall i$. 
The RV $SX_L$ is defined as the sum of the above set, i.e,
\begin{equation}
SX_L = \sum_{i=1}^{L} X_i.
\end{equation}
If $b_i > 0$, $\forall i$, $x \in (sm_L,\infty)$ and if $b_i < 0$, $\forall i$, $x \in (-\infty,sm_L)$ with $sm_L = \sum_{i=1}^{L} m_i$.

\begin{proposition} \label{prop_sum_p}
	The PDF of $SX_L$ is given by
	\begin{equation} \label{pdfSX}
		f_{SX_L}(x) = \begin{cases}
			f_X(x,sa_L,b,sm_L), \qquad \qquad b_i = b, \ \forall i \\
			\begin{split}
				& \sum_{i=1}^{L} \sum_{k=1}^{a_{i}} \Xi_{L}^0 \left(i, k, \left\{a_q\right\}_{q=1}^{L}, \left\{b_q\right\}_{q=1}^{L}, \left\{j_q\right\}_{q=1}^{L-2}\right) \\
				& \quad \times f_{X_i}(x,k,b_i,sm_L),  \qquad b_i \neq b_j, \ \forall i \neq j,
			\end{split} 
		\end{cases}
	\end{equation}
	where the weights $\Xi_{L}^0$ are given by \eqref{Xi0} at the top of the next page with $sa_L = \sum_{i=1}^{L} a_i$ and $U(\cdot)$ being the Heaviside step function defined as $U(x \geq 0) = 1$. 
\end{proposition}

\begin{IEEEproof}
	The proof is provided in Appendix \ref{proof_sum_p}.
\end{IEEEproof}

\begin{figure*}[t]
	\begin{equation} \label{Xi0}
	\begin{split}
	\Xi_{L}^0 & \left(i, k, \left\{a_q\right\}_{q=1}^{L}, \left\{b_q\right\}_{q=1}^{L}, \left\{j_q\right\}_{q=1}^{L-2}\right) 
	= \sum_{j_{1}=k}^{a_{i}} \sum_{j_{2}=k}^{n_{1}} \cdots \sum_{j_{L-2}=k}^{j_{L-3}} 
	(-1)^{sa_L - a_{i}} \frac{\prod_{w=1}^{L} b_{w}^{a_{w}}}{b_{i}^{k}} \frac{\left(a_{i}+a_{1+U(1-i)}-j_{1}-1\right) !}{\left(a_{1+U(1-i)}-1\right) !\left(a_{i}-j_{1}\right) !} \\
	& \times \left(b_i-b_{1+U(1-i)}\right)^{j_{1}-a_{i}-a_{1+U(1-i)}}
	\frac{\left( j_{L-2}+a_{L-1+U(L-1-i)}-k-1 \right) !}{ \left(a_{L-1+U(L-1-i)}-1\right) !\left(j_{L-2}-k\right) ! } 
	\left(b_{i} - b_{L-1+U(L-1-i)}\right)^{k-j_{L-2}-a_{L-1+U(L-1-i)}} \\
	& \times \prod_{s=1}^{L-3} \frac{ \left(j_{s}+a_{s+1+U(s+1-i)} - j_{s+1}-1\right) ! \left( b_{i} - b_{s+1+U(s+1-i)} \right)^{j_{s+1}-j_{s}-a_{s+1+U(s+1-i)}}}{ \left(a_{s+1+U(s+1-i)}-1\right) !\left(j_{s} - j_{s+1}\right) !}
	\end{split}
	\end{equation}	
	\hrule
\end{figure*}

Since \eqref{Xi0} is complicated, a recursive formula for the calculation of $\Xi_{L}^0$ \cite{Coelho1998,Karagiannidis2006} is presented starting from the case that $k=m_i$, i.e,
\begin{equation} \label{Xi0_mi}
	\begin{split}
		\Xi_{L}^0 & \left(i, a_i, \left\{a_q\right\}_{q=1}^{L}, \left\{b_q\right\}_{q=1}^{L}, \left\{j_q\right\}_{q=1}^{L-2}\right) \\
			& \qquad \qquad = \frac{\prod_{w=1}^{L} b_{w}^{a_{w}}}{b_{i}^{a_i}}
			\prod_{\genfrac{}{}{0pt}{}{j=1}{j \neq i}}^{L}\left(b_{j}-b_{i}\right)^{-a_{j}}.
	\end{split}
\end{equation}
When the second argument of $\Xi_{L}^0$ is $a_i-k$ the value of $\Xi_{L}^0$ can be calculated as
\begin{equation} \label{Xi0_mi_k}
	\begin{split}
		\Xi_{L}^0 & \left(i, a_i - k, \left\{a_q\right\}_{q=1}^{L}, \left\{b_q\right\}_{q=1}^{L}, \left\{j_q\right\}_{q=1}^{L-2}\right) \\
			& = \frac{1}{k} \sum_{j=1}^{k} \sum_{\genfrac{}{}{0pt}{}{q=1}{q \neq i}}^{L}  a_{q}b_i^{j} \left(b_{i}-b_{q}\right)^{-j} \\
			& \times \Xi_{L}^0 \left(i, a_i - k + j, \left\{a_q\right\}_{q=1}^{L}, \left\{b_q\right\}_{q=1}^{L}, \left\{j_q\right\}_{q=1}^{L-2}\right).
	\end{split}
\end{equation}

\begin{remark}
	The PDF of $SX_L$ for the case that at least one $m_i \neq 0$ and $b_i \neq b_j$, $\forall i \neq j$ can be written as
	\begin{equation} \label{pdfSX_2}
		\begin{split}
			f_{SX_L}&(x) = \sum_{i=1}^{L} \sum_{k=1}^{a_{i}} \sum_{l=1}^{k}  f_{X_i}(x,l,b_i,m_i) \\
				& \times \Xi_{L} \left(i, k, l, \left\{a_q\right\}_{q=1}^{L}, \left\{b_q\right\}_{q=1}^{L}, \left\{m_q\right\}_{q=1}^{L}, \left\{j_q\right\}_{q=1}^{L-2}\right).
		\end{split}
	\end{equation}
	where the weights $\Xi_{L}$ are given by \eqref{Xi} at the bottom of this page.
	It should be highlighted that the PDF of $SX_L$ in \eqref{pdfSX_2} is a nested finite weighted sum of Pearson type III PDFs.
\end{remark}

\begin{IEEEproof}
	Utilizing the binomial theorem in \eqref{pdfSX}, \eqref{pdfSX_2} can be derived.
\end{IEEEproof}

\begin{figure*}[b]
	\hrule
	\begin{equation} \label{Xi}
	\begin{split}
	\Xi_{L} & \left(i,  k, l, \left\{a_q\right\}_{q=1}^{L}, \left\{b_q\right\}_{q=1}^{L}, \left\{m_q\right\}_{q=1}^{L}, \left\{j_q\right\}_{q=1}^{L-2}\right) 
	= \sum_{j_{1}=k}^{a_{i}} \sum_{j_{2}=k}^{n_{1}} \cdots \sum_{j_{L-2}=k}^{j_{L-3}} 
	\frac{e^{(sm_L - m_i)b_i} (m_i - sm_L)^{k-l}}{(k-l)!} (-1)^{sa_L - a_{i}} \frac{\prod_{w=1}^{L} b_{w}^{a_{w}}}{b_{i}^{l}}  \\
	& \times \frac{\left(a_{i}+a_{1+U(1-i)}-j_{1}-1\right) !}{\left(a_{1+U(1-i)}-1\right) !\left(a_{i}-j_{1}\right) !} \left(b_i-b_{1+U(1-i)}\right)^{j_{1}-a_{i}-a_{1+U(1-i)}}		
	\frac{\left( j_{L-2}+a_{L-1+U(L-1-i)}-k-1 \right) !}{ \left(a_{L-1+U(L-1-i)}-1\right) !\left(j_{L-2}-k\right) ! } \\
	& \times \left(b_{i} - b_{L-1+U(L-1-i)}\right)^{k-j_{L-2}-a_{L-1+U(L-1-i)}} \prod_{s=1}^{L-3} \tfrac{ \left(j_{s}+a_{s+1+U(s+1-i)} - j_{s+1}-1\right) ! \left( b_{i} - b_{s+1+U(s+1-i)} \right)^{j_{s+1}-j_{s}-a_{s+1+U(s+1-i)}}}{ \left(a_{s+1+U(s+1-i)}-1\right) !\left(j_{s} - j_{s+1}\right) !}
	\end{split}
	\end{equation}	
\end{figure*}

In the following proposition, the CDF of the distribution of the sum of Pearson type III RVs is provided.
\begin{proposition}
	The CDF of $SX_L$ is given by 
	\begin{equation} \label{cdfSX}
		F_{SX_L}(x) = \begin{cases}
			F_X(x,sa_L,b,sm_L), \qquad \qquad b_i = b, \ \forall i \\
			\begin{split}
				& \sum_{i=1}^{L} \sum_{k=1}^{a_{i}} \Xi_{L}^0 \left(i, k, \left\{a_q\right\}_{q=1}^{L}, \left\{b_q\right\}_{q=1}^{L}, \left\{j_q\right\}_{q=1}^{L-2}\right) \\
				& \quad \times F_{X_i}(x,k,b_i,sm_L), \qquad b_i \neq b_j, \ \forall i \neq j,
			\end{split} 
		\end{cases}
	\end{equation}
\end{proposition}

\begin{IEEEproof}
	The CDF of $SX_L$ can be obtained by integrating \eqref{pdfSX} from $sm_L$ to $x$, interchanging the order of summations and integrations and utilizing \eqref{PT3_CDF}.
\end{IEEEproof}

In the next proposition, closed-form expression for the moments the distribution of the sum of Pearson type III RVs are provided.
\begin{proposition}
	The $n$-th moment of $SX_L$ is given by 
	\begin{equation} \label{momSX}
		\mu_{SX_L}^n = \begin{cases}
			\mu_X^n (sa_L,b,sm_L), \qquad \quad b_i = b, \ \forall i \\
			\begin{split}
				& \sum_{i=1}^{L} \sum_{k=1}^{a_{i}} \Xi_{L}^0 \left(i, k, \left\{a_q\right\}_{q=1}^{L} \!, \left\{b_q\right\}_{q=1}^{L} \!, \left\{j_q\right\}_{q=1}^{L-2}\right) \\
				& \quad \times \mu_X^n (k,b,sm_L), \ \ \quad b_i \neq b_j, \ \forall i \neq j,
			\end{split}
		\end{cases}
	\end{equation}
\end{proposition}

\begin{IEEEproof}
	The $n$-th moment of $SX_L$ can be obtained by the integrals $\int_{-\infty}^{sm_L} x^n f_{SX_L}(x) dx$ if $b_i<0$, $\forall i$ and $\int_{sm_L}^{\infty} x^n f_{SX_L}(x) dx$ if $b_i>0$, $\forall i$, by interchanging the order of summations and integrations and utilizing \eqref{PT3_moments}.
\end{IEEEproof}

\subsection{Distribution of the Log Sum of Pearson Type III RVs}

In this subsection, the distribution of the log sum of Pearson type III RVs is investigated.
If the RV $SX_L$ is a sum of Pearson type III distributions, the RV $SY_L = e^{SX_L}$ follows the distribution of the log sum of Pearson type III RVs.
If $b_i>0$, $\forall i$, $y\in (e^{sm_L},+\infty)$ and if $b_i<0$, $\forall i$, $y\in (0,e^{sm_L})$.

\begin{proposition}
	The CDF of $SY_L$ is given by 
	\begin{equation} \label{cdfSY}
		F_{SY_L}(y) = \begin{cases}
			F_Y(y,sa_L,b,sm_L), \qquad \qquad b_i = b, \ \forall i \\
			\begin{split}
				& \sum_{i=1}^{L} \sum_{k=1}^{a_{i}} \Xi_{L}^0 \left(i, k, \left\{a_q\right\}_{q=1}^{L}, \left\{b_q\right\}_{q=1}^{L}, 		\left\{j_q\right\}_{q=1}^{L-2}\right) \\
				& \quad \times F_{Y_i}(y,k,b_i,sm_L), \qquad b_i \neq b_j, \ \forall i \neq j.
			\end{split}
		\end{cases}
	\end{equation}
\end{proposition}

\begin{IEEEproof}
	The CDF of $SY_L$ can be obtained from \eqref{cdfSX}.
\end{IEEEproof}

In the following proposition, the PDF of the distribution of the log sum of Pearson type III RVs is extracted.
\begin{proposition}
	The PDF of $SY_L$ is given by 
	\begin{equation} \label{pdfSY}
		f_{SY_L}(y) = \begin{cases}
			f_Y(y,sa_L,b,sm_L), \qquad \qquad  b_i = b, \ \forall i \\
			\begin{split}
				& \sum_{i=1}^{L} \sum_{k=1}^{a_{i}} \Xi_{L}^0 \left(i, k, \left\{a_q\right\}_{q=1}^{L}, \left\{b_q\right\}_{q=1}^{L}, \left\{j_q\right\}_{q=1}^{L-2}\right) \\
				& \quad \times f_{Y_i}(y,k,b_i,sm_L), \qquad b_i \neq b_j, \ \forall i \neq j.
			\end{split}
		\end{cases}
	\end{equation}		
\end{proposition}

\begin{IEEEproof}
	The PDF of $SY_L$ can be obtained as the first derivative of \eqref{cdfSY} and by utilizing \eqref{lP_PDF}.
\end{IEEEproof}

In the next proposition, the moments of the distribution of the log sum of Pearson type III RVs are provided.
\begin{proposition}
	The $n$-th moment of $SY_L$ is given by 
	\begin{equation} \label{momSY}
		\mu_{SY_L}^n = \begin{cases}
			\mu_Y^n (sa_L,b,sm_L), \ \qquad \qquad b_i = b, \ \forall i \\
			\begin{split}
				& \sum_{i=1}^{L} \sum_{k=1}^{a_{i}} \Xi_{L}^0 \left(i, k, \left\{a_q\right\}_{q=1}^{L}, \left\{b_q\right\}_{q=1}^{L}, \left\{j_q\right\}_{q=1}^{L-2}\right) \\
				& \quad \times \mu_Y^n (k,b,sm_L), \qquad \quad b_i \neq b_j, \ \forall i \neq j,
			\end{split}
		\end{cases}
	\end{equation}	
	If $b_i>0$, $\forall i$, $b_i>n$, $\forall i$ should be satisfied.
	If $b_i<0$, $\forall i$, there is no constraint.
\end{proposition}

\begin{IEEEproof}
	The $n$-th moment of $SY_L$ can be obtained by the integrals $\int_{-\infty}^{e^{sm_L}} y^n f_{SY_L}(y) dy$ if $b_i<0$, $\forall i$ and $\int_{e^{sm_L}}^{\infty} y^n f_{SY_L}(y) dy$ if $b_i>0$, $\forall i$, by interchanging the order of summations and integrations and utilizing \eqref{LPT3_moments}.
\end{IEEEproof}

\section{The Logit Pearson Type III Distribution}

In this section, we utilize the derived results  of the previous section to introduce a new member of the Pearson type III family, the logit Pearson Type III distribution, and derive its statistical properties.
In \cite{Johnson1949,Frederic2008}, the logit normal distribution is investigated where, considering that the RV $A$ follows the normal distribution, the RV $B$ follows the logit normal distribution, if $A = \mathrm{logit}(B)=\ln \frac{B}{1-B}$ or $B=f(A)$, where $f(x)=\frac{1}{1+e^{-x}}$ is the logistic function. 
Accordingly, in this work, we introduce the logit Pearson type III distribution, which is defined below.
\begin{definition}
The RV $Z = \frac{1}{1+e^{-X}}$ follows the logit Pearson type III distribution with the parameters  parameters $(a,b,m)$, if the RV $X$ follows the Pearson type III distribution with the same distribution or, equivalently, $X=\mathrm{logit}(Z)$. For the domain of $z$, it holds that $z\in (\frac{1}{1+e^{-m}},1)$ if $b>0$, while $z\in (0,\frac{1}{1+e^{-m}})$ if $b<0$. 
\end{definition}


\begin{proposition}
	The CDF of the logit Pearson type III distribution can be expressed as
	\begin{equation} \label{ltP_CDF}
	F_Z(z,a,b,m) = \begin{cases}
	\frac{1}{\Gamma(a)} \Gamma \left(a, b \left(\ln \frac{z}{1-z} - m\right)\right), &   b < 0 \\
	\frac{1}{\Gamma(a)} \gamma \left(a, b \left(\ln \frac{z}{1-z} - m\right)\right), &   b > 0.
	\end{cases}
	\end{equation}
\end{proposition}

\begin{IEEEproof}
	The CDF of the logit Pearson type III distribution is derived by substituting $x = \ln \frac{z}{1-z}$ in \eqref{PT3_CDF}.
\end{IEEEproof}

In the following proposition, the PDF of $Z$ is extracted.
\begin{proposition}
	The PDF of the logit Pearson type III distribution is given by
	\begin{equation} \label{ltP_PDF}
		\begin{split}
			f_Z(z,a,b,m) & = \frac{|b| e^{bm}}{\Gamma(a)} \left(b\left(\ln \frac{z}{1-z} - m\right)\right)^{a-1} \\
				& \times z^{-b-1} (1-z)^{b-1}.
		\end{split}
	\end{equation}
\end{proposition}

\begin{IEEEproof}
	The PDF of the logit Pearson type III distribution is derived as the first derivative of the CDF given by \eqref{ltP_CDF} and after some algebraic manipulations.
\end{IEEEproof}


\begin{figure}[h]
	\centering
	\begin{tikzpicture}
	\begin{axis}[
	width=0.9\linewidth,
	xlabel = z,
	ylabel = CDF,
	xmin = 0,xmax = 1,
	ymin = 0,ymax = 1,
	grid = major,
	legend style={font=\tiny},
	legend entries = {{$a=3,b=-1.5$},{$a=2,b=-1.5$},{$a=3,b=1.5$},{$a=2,b=1.5$},{Theoretical}},
	legend cell align = {left},
	legend style={at={(0.425,0.95)},anchor=north west}
	]
	\addplot[
	black,
	mark=square,	mark phase = 50,	mark repeat = 50,	mark size = 2,
	only marks
	]
	table {data/cdf/cdf11.dat};
	\addplot[
	black,
	mark=triangle,	mark phase = 50,	mark repeat = 50,	mark size = 2,
	only marks
	]
	table {data/cdf/cdf12.dat};
	\addplot[
	black,
	mark=diamond,	mark phase = 50,	mark repeat = 50,	mark size = 2,
	only marks
	]
	table {data/cdf/cdf21.dat};
	\addplot[
	black,
	mark=triangle,	mark phase = 50,	mark options={rotate=180},	mark repeat = 50,	mark size = 2,
	only marks
	]
	table {data/cdf/cdf22.dat};
	\addplot[
	black,
	no marks,
	line width = 1pt,	style = solid
	]
	table {data/cdf/cdf11.dat};
	\addplot[
	black,
	no marks,
	line width = 1pt,	style = solid
	]
	table {data/cdf/cdf12.dat};
	\addplot[
	black,
	no marks,
	line width = 1pt,	style = solid
	]
	table {data/cdf/cdf21.dat};
	\addplot[
	black,
	no marks,
	line width = 1pt,	style = solid
	]
	table {data/cdf/cdf22.dat};
	\end{axis}
	\end{tikzpicture}
	\caption{CDF of the logit Pearson type III distribution.}
	\label{fig:cdf}
\end{figure}
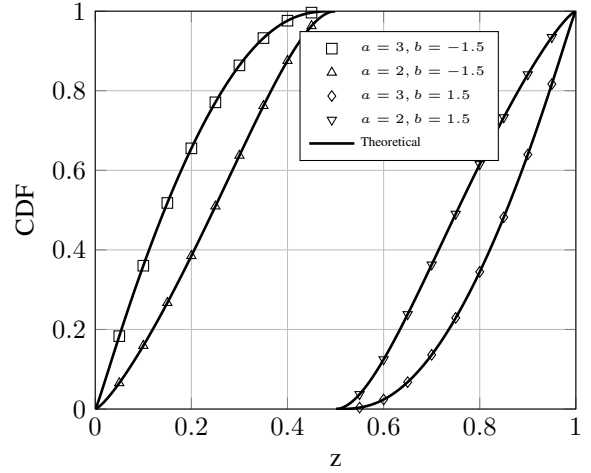

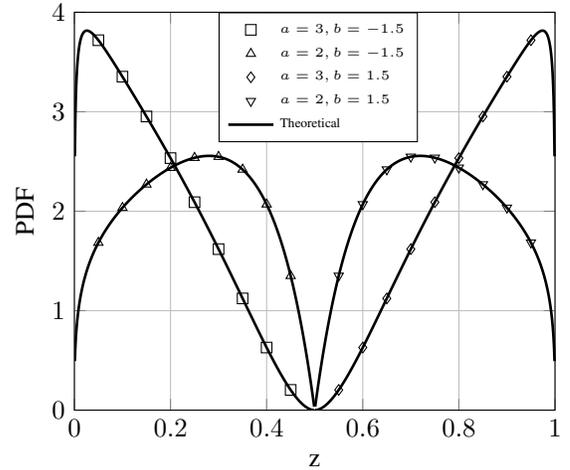
\begin{figure}[h]
	\centering
	\begin{tikzpicture}
	\begin{axis}[
	width=0.9\linewidth,
	xlabel = z,
	ylabel = PDF,
	xmin = 0,xmax = 1,
	ymin = 0,ymax = 4,
	grid = major,
	legend style={font=\tiny},
	legend entries = {{$a=3,b=-1.5$},{$a=2,b=-1.5$},{$a=3,b=1.5$},{$a=2,b=1.5$},{Theoretical}},
	legend cell align = {left},
	legend style={at={(0.3,1)},anchor=north west}
	]
	\addplot[
	black,
	mark=square,	mark phase = 50,	mark repeat = 50,	mark size = 2,
	only marks
	]
	table {data/pdf/pdf11.dat};
	\addplot[
	black,
	mark=triangle,	mark phase = 50,	mark repeat = 50,	mark size = 2,
	only marks
	]
	table {data/pdf/pdf12.dat};
	\addplot[
	black,
	mark=diamond,	mark phase = 50,	mark repeat = 50,	mark size = 2,
	only marks
	]
	table {data/pdf/pdf21.dat};
	\addplot[
	black,
	mark=triangle,	mark phase = 50,	mark options={rotate=180},	mark repeat = 50,	mark size = 2,
	only marks
	]
	table {data/pdf/pdf22.dat};
	\addplot[
	black,
	no marks,
	line width = 1pt,	style = solid
	]
	table {data/pdf/pdf11.dat};
	\addplot[
	black,
	no marks,
	line width = 1pt,	style = solid
	]
	table {data/pdf/pdf12.dat};
	\addplot[
	black,
	no marks,
	line width = 1pt,	style = solid
	]
	table {data/pdf/pdf21.dat};
	\addplot[
	black,
	no marks,
	line width = 1pt,	style = solid
	]
	table {data/pdf/pdf22.dat};
	\end{axis}
	\end{tikzpicture}
	\caption{PDF of the logit Pearson type III distribution.}
	\label{fig:pdf}
\end{figure}
In Figs. \ref{fig:cdf} and \ref{fig:pdf}, the CDF and the PDF of the introduced logit Pearson type III distribution are illustrated, respectively. 
In both Figs., we set $m=0$, thus for negative $b$, $x\in(0,0.5)$ and for positive $b$, $x\in(0.5,1)$. It should be highlighted that neither PDF nor CDF is defined in $0$, $0.5$ or $1$ and for the case that $m=0$, the PDF is symmetric around $0.5$. It can be observed that the simulations validate the theoretical results.

\begin{proposition} \label{prop_lpt_mom}
	The $n$-th moment of the logit Pearson type III distribution when $b>0$ is given by \eqref{ltPT3_moments} at the top of the next page.
\end{proposition}

\begin{figure*}[t]
	\begin{equation} \label{ltPT3_moments}
		\mu_Z^n (a,b,m) = \begin{cases}
			\sum_{l=0}^\infty \binom{n+l-1}{l} (-1)^l e^{-m l} \left(1+\frac{l}{b}\right)^{-a}, \qquad \qquad \qquad \qquad m \geq 0 \\
			\begin{split}
				& \frac{1}{\Gamma(a)} \sum_{l=0}^\infty \binom{n+l-1}{l} (-1)^l \left(e^{m (n+l)} \left(1-\frac{n+l}{b}\right)^{-a} \gamma\left(a,-m b \left(1-\frac{n+l}{b}\right)\right) \right. \\
				& \left. \qquad + e^{-m l} \left(1+\frac{l}{b}\right)^{-a} \Gamma\left(a,-m b \left(1+\frac{l}{b}\right)\right) \right), \ \ \qquad m < 0
			\end{split} 
		\end{cases}
	\end{equation}
\hrule
\end{figure*}

\begin{IEEEproof}
	The proof is provided in Appendix \ref{proof_lpt_mom}.
\end{IEEEproof}

\begin{corollary} \label{m1}
	The mean value of the logit Pearson type III distribution when $b>0$ and $m \geq 0$ is given in closed-form by
	\begin{equation}
	\mu_Z^1 (a,b,m) = b^a \Phi \left( -e^{-m} , a , b \right),
	\end{equation}
	where $\Phi(\cdot,\cdot,\cdot)$ is the Lerch function \cite{Gradshteyn2014}.
\end{corollary}
\begin{corollary} \label{m2}
	The second moment of the logit Pearson type III distribution when $b>0$ and $m \geq 0$ is given in closed-form by
	\begin{equation}
		\begin{split}
			\mu_Z^2 (a,b,m) & = b^a \left( \Phi \left( -e^{-m} , a-1 , b \right) \right. \\
				& \left. - (b-1) \Phi \left( -e^{-m} , a , b \right) \right).
		\end{split}
	\end{equation}
\end{corollary}

\subsection{The Logit Gamma Distribution}

Utilizing the above analysis, for the special case that $b>0$ and $m=0$, the logit gamma distribution can be derived. If the RV $Z$ follows the logit gamma distribution, it holds that $z\in(0.5,1)$.
The CDF of the logit gamma distribution can be expressed as
\begin{equation} \label{ltG_CDF}
	F_Z(z,a,b) = \frac{1}{\Gamma(a)} \gamma \left(a, b \ln \frac{z}{1-z} \right).
\end{equation}
The PDF of the logit gamma distribution is given by
\begin{equation} \label{ltG_PDF}
	f_Z(z,a,b) = \frac{b}{\Gamma(a)} \left(b \ln \frac{z}{1-z} \right)^{a-1} z^{-b-1} (1-z)^{b-1}.
\end{equation}
The $n$-th moment of the logit gamma distribution is given by 
\begin{equation} \label{ltG_moments}
\mu_Z^n (a,b) = \sum_{l=0}^\infty \binom{n+l-1}{l} (-1)^l \left(1+\frac{l}{b}\right)^{-a}.
\end{equation}
Corollaries \ref{m1} and \ref{m2} can be used to extract closed-form expressions for the first and the second moment of the logit gamma distribution.

\subsection{Distribution of the Logit Sum of Pearson Type III RVs}

In this subsection, the distribution of the logit sum of Pearson type III RVs is investigated.
If the RV $SX_L$ is a sum of Pearson type III RVs, the RV $SZ_L = \frac{1}{1+e^{-SX_L}}$ follows the distribution of the logit sum of Pearson type III RVs.
If $b_i>0$, $\forall i$, $z\in (\frac{1}{1+e^{-sm_L}},1)$ and if $b_i<0$, $\forall i$, $z\in (0,\frac{1}{1+e^{-sm_L}})$.

\begin{proposition}
	The CDF of $SZ_L$ is given by 
	\begin{equation} \label{cdfSZ}
		F_{SZ_L}(z) = \begin{cases}
			F_Z(z,sa_L,b,sm_L), \qquad \qquad b_i = b, \ \forall i \\
			\begin{split}
				& \sum_{i=1}^{L} \sum_{k=1}^{a_{i}} \Xi_{L}^0 \left(i, k, \left\{a_q\right\}_{q=1}^{L}, \left\{b_q\right\}_{q=1}^{L}, \left\{j_q\right\}_{q=1}^{L-2}\right) \\
				& \quad \times F_{Z_i}(z,k,b_i,sm_L), \qquad b_i \neq b_j, \ \forall i \neq j.
			\end{split}
		\end{cases}
	\end{equation}
\end{proposition}

\begin{IEEEproof}
	The CDF of $SZ_L$ can be obtained from \eqref{cdfSX} by substituting $x = \ln \frac{z}{1-z}$.
\end{IEEEproof}

In the following proposition, the PDF of the distribution of the logit sum of Pearson type III RVs is extracted.
\begin{proposition}
	The PDF of $SZ_L$ is given by 
	\begin{equation} \label{pdfSZ}
		f_{SZ_L}(z) = \begin{cases}
			f_Z(z,sa_L,b,sm_L), \qquad \qquad b_i = b, \ \forall i \\
			\begin{split}
				& \sum_{i=1}^{L} \sum_{k=1}^{a_{i}} \Xi_{L}^0 \left(i, k, \left\{a_q\right\}_{q=1}^{L}, \left\{b_q\right\}_{q=1}^{L}, \left\{j_q\right\}_{q=1}^{L-2}\right) \\
				& \quad \times f_{Z_i}(z,k,b_i,sm_L), \qquad b_i \neq b_j, \ \forall i \neq j.
			\end{split}
		\end{cases}
	\end{equation}	
\end{proposition}

\begin{IEEEproof}
	The PDF of $SZ_L$ can be obtained as the first derivative of \eqref{cdfSZ} and by utilizing \eqref{ltP_PDF}.
\end{IEEEproof}

In the next proposition, the moments of the distribution of the logit sum of Pearson type III RVs are provided.
\begin{proposition}
	The $n$-th moment of $SZ_L$ when $b_i > 0, \ \forall i$ is given by 
	\begin{equation} \label{momSZ}
		\mu_{SZ_L}^n = \begin{cases}
			\mu_Z^n (sa_L,b,sm_L), \qquad \qquad \ b_i = b, \ \forall i \\
			\begin{split}
				& \sum_{i=1}^{L} \sum_{k=1}^{a_{i}} \Xi_{L}^0 \left(i, k, \left\{a_q\right\}_{q=1}^{L}, \left\{b_q\right\}_{q=1}^{L}, \left\{j_q\right\}_{q=1}^{L-2}\right) \\
				& \quad \times \mu_Z^n (k,b,sm_L), \qquad \quad b_i \neq b_j, \ \forall i \neq j,
			\end{split}
		\end{cases}
	\end{equation}	
\end{proposition}

\begin{IEEEproof}
	The $n$-th moment of $SZ_L$ can be obtained by the integral $\int_{\frac{1}{1+e^{-sm_L}}}^{1} z^n f_{SZ_L}(z) dz$ by interchanging the order of summations and integrations and utilizing \eqref{ltPT3_moments}.
\end{IEEEproof}

\section{Application of Pearson Type III Family of Distributions in WPT}

\subsection{System Model}

In this section, a network is considered which consists of one PB or multiple PBs that utilizes WPT to provide energy to the assigned EH sources. It is assumed that the harvested power due to the processing noise is negligible and, thus, it can be ignored. 
The nonlinear EH model proposed in \cite{Boshkovska2015} is considered.
In contrast with the linear EH model which is accurate only when the received power is constant, this nonlinear model captures the dynamics of the RF energy conversion efficiency for different input power levels and is based on the logistic function.
Two scenarios are considered, i.e., a SISO scenario and a MISO one. Also, for the MISO scenario two cases are considered, i.e., a network with a PB with $L$ antennas and a network with $L$ PBs with a single antenna. 
When multiple antennas in the PB or multiple PBs are considered, the assigned EH sources can harvest energy more reliably than the SISO case, since the values of the harvested energy, that are lower than the sensitivity threshold, which results from the harvesting circuits, due to the randomness of the fading, reduce. Therefore, the harvested energy increases for the MISO scenario.

The power harvested by one of the sources for the SISO scenario can be expressed as \cite{Boshkovska2015}
\begin{equation} \label{Q_nl_SISO}
Q^S = \frac{P_s\left(1+e^{AB}\right)}{e^{AB}\left(1+e^{-A(l p |h|^2 -B)}\right)}-\frac{P_s}{e^{AB}},
\end{equation}
where $P_s$ denotes the maximum harvested power when the EH circuit is saturated. Also, $A$ and $B$ are positive constants related to the circuit specification. Practically, $A$ reflects the nonlinear charging rate with respect to the input power and $B$ is related to the turn-on threshold. Given the EH circuit, the parameters $P_s$, $A$, and $B$ can be determined by the curve fitting. 
Furthermore, $l$, $p$ and $h$ denote the path loss factor between the PB and the source, the transmitted power and the small scale fading coefficient between the PB and the source, respectively.
We assume that the channel fading between the PB and the source is a stationary and ergodic random process, whose instantaneous channel realizations follow the Nakagami distribution with parameters $(a,\frac{a}{b})$, since the Nakagami channel model is general enough to describe the typical wireless fading environments.
In this case, $|h|^2$ follows the gamma distribution with parameters $(a,b)$ or the Pearson type III distribution with parameters $(a,b,0)$.

The power harvested by one of the sources for the MISO scenario can be expressed as
\begin{equation} \label{Q_nl_MISO}
Q^M = \frac{P_s\left(1+e^{AB}\right)}{e^{AB}\left(1+e^{-A \left( \sum_{i=1}^L l_i p_i |h_i|^2 -B \right)  }\right)}-\frac{P_s}{e^{AB}},
\end{equation}
where $L$ denotes the number of antennas. It should be highlighted that the sum of Pearson type III distributions appears, thus the distributions presented in Section III can be utilized. In the case that the network consists of one PB with $L$ antennas, assuming that the available power is equally split into the $L$ antennas, it holds that $\frac{b_i}{l_i p_i} = \frac{b}{l p}, \ \forall i$. In the second case where the network consists of $L$ PBs with one antenna, assuming that the distance between the EH source and each PB is different, it holds that $\frac{b_i}{l_i p_i} \neq \frac{b_j}{l_j p_j}, \ \forall i \neq j$.

\subsection{Statistical Properties}

\subsubsection{SISO} 
Some important statistical properties of the distribution of the harvested power for the SISO case are presented below.

\begin{theorem}
	The CDF of the distribution of the harvested power for the SISO case is given by
	\begin{equation} \label{EH_CDF}
		\begin{split}
			F_{Q^S}(q) & = \frac{1}{\Gamma(a)} \gamma \left( a , - \frac{b}{A l p} \right. \\
				& \left. \times \left( \ln \left(\frac{P_s\left(1+e^{AB}\right)}{e^{AB}\left(q + \frac{P_s}{e^{AB}}\right)} - 1 \right) - A B \right) \right).
		\end{split}
	\end{equation}
\end{theorem}

\begin{IEEEproof}
	In \eqref{Q_nl_SISO}, $|h|^2$ follows the gamma distribution with parameters $(a,b)$, which is also the Pearson type III distribution with parameters $(a,b,0)$ and $l p |h|^2$ follows the Pearson type III distribution with parameters $(a,\frac{b}{l p},0)$.
	Therefore, $-A\left(l p |h|^2 - B\right)$ follows the Pearson type III distribution with parameters $(a, -\frac{b}{A l p}, A B)$, $e^{-A\left(l p |h|^2 - B\right)}$ follows the log Pearson type III distribution with the same parameters and $\frac{1}{1+e^{-A\left(l p |h|^2 - B\right)}}$ follows the logit Pearson type III distribution with parameters $(a, \frac{b}{A l p}, -A B)$.
	
	The CDF of the distribution of the harvested power is obtained as
	\begin{equation} \label{pr1}
	F_{Q^S}(q) = P \left( Q < q \right),
	\end{equation}
	where $P(\cdot)$ denotes probability.
	After some algebraic manipulations and using \eqref{ltP_CDF}, \eqref{pr1} can be rewritten as
	\begin{equation} \label{EH_cdf_proof}
	F_{Q^S}(q) = F_Z \left( \frac{e^{AB}\left(q + \frac{P_s}{e^{AB}}\right)}{P_s\left(1+e^{AB}\right)} \right).
	\end{equation} 
	From \eqref{EH_cdf_proof}, \eqref{EH_CDF} is derived.
\end{IEEEproof}
It should be highlighted that the CDF of the distribution of the harvested power indicated the probability that outage occurs in the harvested power if we consider a threshold $q$.

In the following theorem, the PDF of $Q^S$ is extracted.
\begin{theorem}
	The PDF of the distribution of the harvested power for the SISO case can be expressed as 
	\begin{equation} \label{EH_PDF}
		\begin{split}
			f_{Q^S}(q) & = \frac{c \left(1+e^{AB}\right) \hat{b}^a}{\Gamma(a) e^{A B \hat{b}}} \left( c e^{AB} - q \right)^{-1+\hat{b}} (c + q)^{-1-\hat{b}} \\
				& \times \left( AB - \ln \left( \frac{c\left(1+e^{AB}\right)}{c+q} - 1 \right) \right)^{a-1},
		\end{split}
	\end{equation}
	where $\hat{b} = \frac{b}{A l p}$ and $c = \frac{P_s}{e^{AB}}$.
\end{theorem}

\begin{IEEEproof}
	The PDF is obtained as the first derivative of the CDF given by \eqref{EH_CDF} and after some algebraic manipulations.
\end{IEEEproof}

In the following theorem, the moments of $Q^S$ are provided.
\begin{theorem} \label{th_EH_moments}
	The $n$-th moment of the distribution of the harvested power for the SISO case is given by \eqref{EH_moments} at the top of the next page,
	\begin{figure*}[t]
		\begin{equation} \label{EH_moments}
		\begin{split}
		\mu_{Q^S}^n & = \frac{c^n}{\Gamma(a)} \sum_{l_1=0}^n \sum_{l_2=0}^\infty \binom{n}{l_1} \binom{l_1+l_2-1}{l_2} (-1)^{n-l_1+l_2} \left(e^{-A B}+1\right)^{l_1}
		\left( e^{-A B l_2} 
		\left(1 - \frac{l_1+l_2}{\hat{b}}\right)^{-a} \right. \\
		& \left. \times \gamma \left(a, A B \hat{b} \left(1 - \frac{l_1+l_2}{\hat{b}}\right)\right) 
		+ e^{A B (l_1+l_2)} \left(1+\frac{l_2}{\hat{b}}\right)^{-a} 
		\Gamma \left(a, A B \hat{b} \left(1+\frac{l_2}{\hat{b}}\right)\right) \right)
		\end{split}
		\end{equation}
		\hrule
	\end{figure*}
	if $\hat{b} \notin \mathbb{Z}$.
\end{theorem}

\begin{IEEEproof}
	The proof is provided in Appendix \ref{proof_EH_moments}.
\end{IEEEproof}

\begin{corollary}
	The mean value is obtained as the first moment and can be expressed as in \eqref{EH_mean} at the top of the next page,
	\begin{figure*}[t]
		\begin{equation} \label{EH_mean}
		\begin{split}
		\mu_{Q^S}^1 & = \frac{c \left(1+e^{-AB}\right)}{\Gamma(a)} \sum _{k=0}^\infty (-1)^k e^{-A B k} \left( \left(-\frac{k+1}{\hat{b}} + 1\right)^{-a} 
		\gamma \left(a,A B \hat{b} \left(-\frac{k+1}{\hat{b}} + 1\right)\right) \right. \\
		& \left. + e^{A B (1+2k)} \left(\frac{k}{\hat{b}} + 1\right)^{-a} 
		\Gamma \left(a,A B \hat{b} \left(\frac{k}{\hat{b}} + 1\right)\right) \right) - c
		\end{split}
		\end{equation}
		\hrule
	\end{figure*}
	if $\hat{b} \notin \mathbb{Z}$.
\end{corollary}

\begin{remark}
	From the second moment the variance of the harvested power can be derived which expresses how the values of the harvested power fluctuate around the mean value. It should be highlighted that the variance should be small, the majority of the harvested power is larger than the sensitivity threshold.
\end{remark}

\subsubsection{MISO}
Accordingly, the statistical properties of the distribution of the harvested power for the MISO case are presented below.

\begin{theorem}
	The CDF of the distribution of the harvested power for the MISO case is given by 
	\begin{equation} \label{EH_CDF_M}
		F_{Q^M}(q) = \begin{cases}
			\begin{split}
				& \frac{1}{\Gamma(sa_L)} \gamma \left( sa_L , - \frac{b_i}{A l_i p_i} \right. \\
				& \ \left. \times \left( \ln \left(\frac{P_s\left(1+e^{AB}\right)}{e^{AB}\left(q + \frac{P_s}{e^{AB}}\right)} - 1 \right) - A B \right) \right), \\
				& \qquad \qquad \frac{b_i}{l_i p_i} = \frac{b}{l p}, \ \forall i 
			\end{split} \\
			\begin{split}
				& \sum_{i=1}^{L} \sum_{k=1}^{a_{i}} \frac{1}{\Gamma(k)} \gamma \left( k , - \frac{b_i}{A l_i p_i} \right. \\
				& \ \left. \times \left( \ln \left(\frac{P_s\left(1+e^{AB}\right)}{e^{AB}\left(q + \frac{P_s}{e^{AB}}\right)} - 1 \right) - A B \right) \right) \\
				& \ \times \Xi_{L}^0 \left(i, k, \left\{a_q\right\}_{q=1}^{L}, \left\{b_q\right\}_{q=1}^{L}, \left\{j_q\right\}_{q=1}^{L-2}\right), \\
				& \qquad \qquad \frac{b_i}{l_i p_i} \neq \frac{b_j}{l_j p_j}, \ \forall i \neq j. 
			\end{split}
		\end{cases}
	\end{equation}
\end{theorem}

\begin{IEEEproof}
	The CDF of the distribution of the harvested power for the MISO case is obtained considering that $\sum_{i=1}^L l_i p_i |h_i|^2$ is a sum of Pearson type III distributions and utilizing \eqref{cdfSX} and \eqref{EH_CDF}.
\end{IEEEproof}

In the next theorem, the PDF of $Q^M$ is extracted.
\begin{theorem}
	The PDF of the distribution of the harvested power for the MISO case can be expressed as 
	\begin{equation} \label{EH_PDF_M}
	f_{Q^M}(q) = \begin{cases}
	\begin{split}
	& \tfrac{c \left(1+e^{AB}\right) \hat{b}_i^{sa_L}}{\Gamma(sa_L) e^{A B \hat{b}_i}} \left( c e^{AB} - q \right)^{-1+\hat{b}_i} (c + q)^{-1-\hat{b}_i} \\
	& \ \times \left( AB - \ln \left( \frac{c\left(1+e^{AB}\right)}{c+q} - 1 \right) \right)^{sa_L-1}, \\
	& \qquad \qquad \frac{b_i}{l_i p_i} = \frac{b}{l p}, \ \forall i 
	\end{split} \\
	\begin{split}
	& \sum_{i=1}^{L} \sum_{k=1}^{a_{i}} \Xi_{L}^0 \left(i, k, \left\{a_q\right\}_{q=1}^{L}, \left\{b_q\right\}_{q=1}^{L}, \left\{j_q\right\}_{q=1}^{L-2}\right) \\
	& \ \times \tfrac{c \left(1+e^{AB}\right) \hat{b}_i^k}{\Gamma(k) e^{A B \hat{b}_i}} \left( c e^{AB} - q \right)^{-1+\hat{b}_i} (c + q)^{-1-\hat{b}_i} \\
	& \ \times \left( AB - \ln \left( \frac{c\left(1+e^{AB}\right)}{c+q} - 1 \right) \right)^{k-1},\\
	& \qquad \qquad \frac{b_i}{l_i p_i} \neq \frac{b_j}{l_j p_j}, \ \forall i \neq j. 
	\end{split}
	\end{cases}
	\end{equation}
\end{theorem}

\begin{IEEEproof}
	The PDF is obtained as the first derivative of the CDF given by \eqref{EH_CDF_M} and after some algebraic manipulations.
\end{IEEEproof}

In the following theorem, the moments of $Q^M$ are provided.
\begin{theorem}
	The $n$-th moment of the distribution of the harvested power for the MISO case is given by \eqref{EH_moments_M} at the top of the next page,
	\begin{figure*}[t]
		\begin{equation} \label{EH_moments_M}
		\mu_{Q^M}^n = \begin{cases}
		\begin{split}
		& \frac{c^n}{\Gamma(sa_L)} \sum_{l_1=0}^n \sum_{l_2=0}^\infty \binom{n}{l_1} \binom{l_1+l_2-1}{l_2} (-1)^{n-l_1+l_2} \left(e^{-A B}+1\right)^{l_1}
		\left( e^{-A B l_2} 
		\left(1 - \frac{l_1+l_2}{\hat{b}}\right)^{-a} \right. \\
		& \quad \left. \times \gamma \left(sa_L, A B \hat{b} \left(1 - \frac{l_1+l_2}{\hat{b}}\right)\right) 
		+ e^{A B (l_1+l_2)} \left(1+\frac{l_2}{\hat{b}}\right)^{-a} 
		\Gamma \left(sa_L, A B \hat{b} \left(1+\frac{l_2}{\hat{b}}\right)\right) \right), \qquad \frac{b_i}{l_i p_i} = \frac{b}{l p},\ \forall i 
		\end{split} \\
		\begin{split}
		& \sum_{i=1}^{L} \sum_{k=1}^{a_{i}} \sum_{l_1=0}^n \sum_{l_2=0}^\infty 
		\Xi_{L}^0 \left(i, k, \left\{a_q\right\}_{q=1}^{L}, \left\{b_q\right\}_{q=1}^{L}, \left\{j_q\right\}_{q=1}^{L-2}\right)
		\frac{c^n}{\Gamma(k)} \binom{n}{l_1} \binom{l_1+l_2-1}{l_2} (-1)^{n-l_1+l_2} \left(e^{-A B}+1\right)^{l_1} \\
		& \quad \left( e^{-A B l_2} 
		\left(1 - \frac{l_1+l_2}{\hat{b}_i}\right)^{-k} \gamma \left(k, A B \hat{b}_i \left(1 - \frac{l_1+l_2}{\hat{b}_i}\right)\right) 
		+ e^{A B (l_1+l_2)} \left(1+\frac{l_2}{\hat{b}_i}\right)^{-k} 
		\Gamma \left(k, A B \hat{b}_i \left(1+\frac{l_2}{\hat{b}_i}\right)\right) \right), \\
		& \ \qquad \qquad \qquad \qquad \qquad \qquad \qquad \qquad \qquad \qquad \qquad \qquad \qquad \qquad \qquad \qquad \qquad \qquad \frac{b_i}{l_i p_i} \neq \frac{b_j}{l_j p_j}, \ \forall i \neq j
		\end{split}
		\end{cases}
		\end{equation}
		\hrule
	\end{figure*}
	if $\hat{b}_i \notin \mathbb{Z}$.
\end{theorem}

\begin{IEEEproof}
	The $n$-th moment is obtained by utilizing \eqref{EH_moments} and \eqref{EH_PDF_M}. 
\end{IEEEproof}

\subsection{Simulation Results}

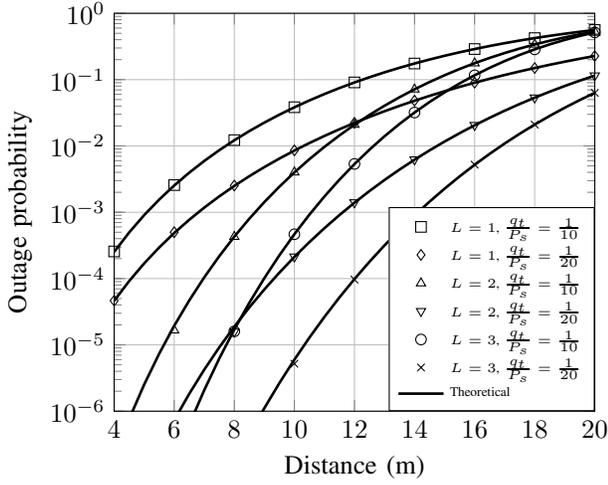
\begin{figure}[h]
	\centering
	\begin{tikzpicture}
	\begin{semilogyaxis}[
	width=0.9\linewidth,
	xlabel = Distance (m),
	ylabel = Outage probability,
	xmin = 4,xmax = 20,
	ymin = 0.000001,ymax = 1,
	grid = major,
	legend style={font=\tiny},
	legend entries = {{$L=1$, $\frac{q_t}{P_s} = \frac{1}{10}$},{$L=1$, $\frac{q_t}{P_s} = \frac{1}{20}$},{$L=2$, $\frac{q_t}{P_s} = \frac{1}{10}$},{$L=2$, $\frac{q_t}{P_s} = \frac{1}{20}$},{$L=3$, $\frac{q_t}{P_s} = \frac{1}{10}$},{$L=3$, $\frac{q_t}{P_s} = \frac{1}{20}$},{Theoretical}},
	legend cell align = {left},
	legend style={at={(axis cs:20,0.000001)},anchor=south east}
	]
	\addplot[
	black,
	mark=square,	mark repeat = 4,	mark size = 2,
	only marks
	]
	table {data/outage_vs_distance/s1_1_8.dat};
	\addplot[
	black,
	mark=diamond,	mark repeat = 4,	mark size = 2,
	only marks
	]
	table {data/outage_vs_distance/s1_1_2.dat};
	\addplot[
	black,
	mark=triangle,	mark phase = 2,		mark repeat = 4,	mark size = 2,
	only marks
	]
	table {data/outage_vs_distance/s2_1_8.dat};
	\addplot[
	black,
	mark=triangle,	mark phase = 3,	mark options={rotate=180},	mark repeat = 4,	mark size = 2,
	only marks
	]
	table {data/outage_vs_distance/s2_1_2.dat};
	\addplot[
	black,
	mark=o,	mark phase = 3,		mark repeat = 4,	mark size = 2,
	only marks
	]
	table {data/outage_vs_distance/s3_1_8.dat};
	\addplot[
	black,
	mark=x,	mark phase = 2,		mark repeat = 4,	mark size = 2,
	only marks
	]
	table {data/outage_vs_distance/s3_1_2.dat};
	\addplot[
	black,
	no marks,
	line width = 1pt,	style = solid
	]
	table {data/outage_vs_distance/t1_1_8.dat};
	\addplot[
	black,
	no marks,
	line width = 1pt,	style = solid
	]
	table {data/outage_vs_distance/t1_1_2.dat};
	\addplot[
	black,
	no marks,
	line width = 1pt,	style = solid
	]
	table {data/outage_vs_distance/t2_1_8.dat};
	\addplot[
	black,
	no marks,
	line width = 1pt,	style = solid
	]
	table {data/outage_vs_distance/t2_1_2.dat};
	\addplot[
	black,
	no marks,
	line width = 1pt,	style = solid
	]
	table {data/outage_vs_distance/t3_1_8.dat};
	\addplot[
	black,
	no marks,
	line width = 1pt,	style = solid
	]
	table {data/outage_vs_distance/t3_1_2.dat};
	\end{semilogyaxis}
	\end{tikzpicture}
	\caption{Outage probability versus distance between the PB and the EH source.}
	\label{fig:ovsd}
\end{figure}

\begin{figure}[h]
	\centering
	\begin{tikzpicture}
	\begin{semilogyaxis}[
	width=0.9\linewidth,
	xlabel = Available transmitted power (W),
	ylabel = Outage probability,
	xmin = 0.5,xmax = 4,
	ymin = 0.000001,ymax = 1,
	grid = major,
	legend style={font=\tiny},
	legend entries = {{$L=1$, $\frac{q_t}{P_s} = \frac{1}{10}$},{$L=1$, $\frac{q_t}{P_s} = \frac{1}{20}$},{$L=2$, $\frac{q_t}{P_s} = \frac{1}{10}$},{$L=2$, $\frac{q_t}{P_s} = \frac{1}{20}$},{$L=3$, $\frac{q_t}{P_s} = \frac{1}{10}$},{$L=3$, $\frac{q_t}{P_s} = \frac{1}{20}$},{Theoretical}},
	legend cell align = {left},
	legend style={at={(axis cs:0.5,0.000001)},anchor=south west}
	]
	\addplot[
	black,
	mark=square,	mark repeat = 2,	mark size = 2,
	only marks
	]
	table {data/outage_vs_power/s1_1_8.dat};
	\addplot[
	black,
	mark=diamond,	mark repeat = 2,	mark size = 2,
	only marks
	]
	table {data/outage_vs_power/s1_1_2.dat};
	\addplot[
	black,
	mark=triangle,	mark repeat = 2,	mark size = 2,
	only marks
	]
	table {data/outage_vs_power/s2_1_8.dat};
	\addplot[
	black,
	mark=triangle,	mark options={rotate=180},	mark repeat = 2,	mark size = 2,
	only marks
	]
	table {data/outage_vs_power/s2_1_2.dat};
	\addplot[
	black,
	mark=o,	mark repeat = 2,	mark size = 2,
	only marks
	]
	table {data/outage_vs_power/s3_1_8.dat};
	\addplot[
	black,
	mark=x,	mark repeat = 2,	mark size = 2,
	only marks
	]
	table {data/outage_vs_power/s3_1_2.dat};
	\addplot[
	black,
	no marks,
	line width = 1pt,	style = solid
	]
	table {data/outage_vs_power/t1_1_8.dat};
	\addplot[
	black,
	no marks,
	line width = 1pt,	style = solid
	]
	table {data/outage_vs_power/t1_1_2.dat};
	\addplot[
	black,
	no marks,
	line width = 1pt,	style = solid
	]
	table {data/outage_vs_power/t2_1_8.dat};
	\addplot[
	black,
	no marks,
	line width = 1pt,	style = solid
	]
	table {data/outage_vs_power/t2_1_2.dat};
	\addplot[
	black,
	no marks,
	line width = 1pt,	style = solid
	]
	table {data/outage_vs_power/t3_1_8.dat};
	\addplot[
	black,
	no marks,
	line width = 1pt,	style = solid
	]
	table {data/outage_vs_power/t3_1_2.dat};
	\end{semilogyaxis}
	\end{tikzpicture}
	\caption{Outage probability versus available transmitted power.}
	\label{fig:ovsp}
\end{figure}
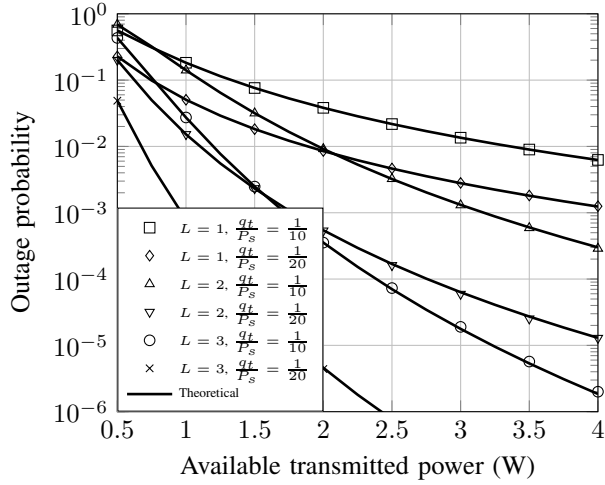

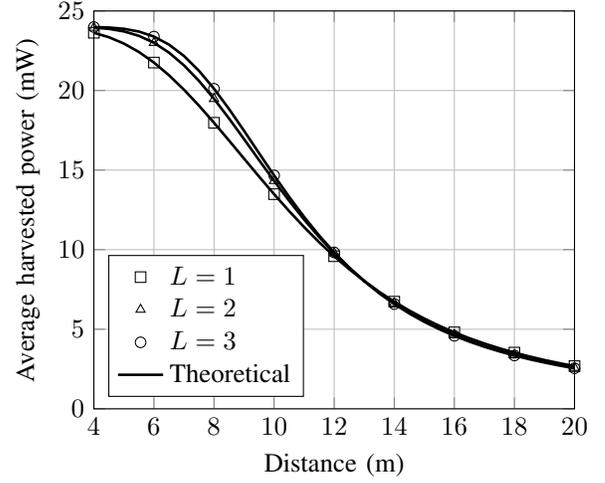
\begin{figure}[h]
	\centering
	\begin{tikzpicture}
	\begin{axis}[
	width=0.9\linewidth,
	xlabel = Distance (m),
	ylabel = Average harvested power (mW),
	xmin = 4,xmax = 20,
	ymin = 0,ymax = 25,
	xtick = {2,4,6,8,10,12,14,16,18,20,22,24},
	grid = major,
	legend entries = {{$L=1$},{$L=2$},{$L=3$},{Theoretical}},
	legend cell align = {left},
	legend pos = south west
	]
	\addplot[
	black,
	mark=square,	mark repeat = 4,	mark size = 2,
	only marks
	]
	table {data/mean_vs_distance/s1.dat};
	\addplot[
	black,
	mark=triangle,	mark repeat = 4,	mark size = 2,
	only marks
	]
	table {data/mean_vs_distance/s2.dat};
	\addplot[
	black,
	mark=o,	mark repeat = 4,	mark size = 2,
	only marks
	]
	table {data/mean_vs_distance/s3.dat};
	\addplot[
	black,
	no marks,
	line width = 1pt,	style = solid
	]
	table {data/mean_vs_distance/s1.dat};
	\addplot[
	black,
	no marks,
	line width = 1pt,	style = solid
	]
	table {data/mean_vs_distance/s2.dat};
	\addplot[
	black,
	no marks,
	line width = 1pt,	style = solid
	]
	table {data/mean_vs_distance/s3.dat};
	\end{axis}
	\end{tikzpicture}
	\caption{Average harvested power versus distance between the PB and the EH source.}
	\label{fig:ahevsd}
\end{figure}

\begin{figure}[h]
	\centering
	\begin{tikzpicture}
	\begin{axis}[
	width=0.9\linewidth,
	xlabel = Available transmitted power (W),
	ylabel = Average harvested power (mW),
	xmin = 0.5,xmax = 4,
	ymin = 0,ymax = 25,
	grid = major,
	legend entries = {{$L=1$},{$L=2$},{$L=3$},{Theoretical}},
	legend cell align = {left},
	legend pos = south east
	]
	\addplot[
	black,
	mark=square,	mark repeat = 5,	mark size = 2,
	only marks
	]
	table {data/mean_vs_power/s1.dat};
	\addplot[
	black,
	mark=triangle,	mark repeat = 5,	mark size = 2,
	only marks
	]
	table {data/mean_vs_power/s2.dat};
	\addplot[
	black,
	mark=o,	mark repeat = 5,	mark size = 2,
	only marks
	]
	table {data/mean_vs_power/s3.dat};
	\addplot[
	black,
	no marks,
	line width = 1pt,	style = solid
	]
	table {data/mean_vs_power/t1.dat};
	\addplot[
	black,
	no marks,
	line width = 1pt,	style = solid
	]
	table {data/mean_vs_power/t2.dat};
	\addplot[
	black,
	no marks,
	line width = 1pt,	style = solid
	]
	table {data/mean_vs_power/t3.dat};
	\end{axis}
	\end{tikzpicture}
	\caption{Average harvested power versus available transmitted power.}
	\label{fig:ahevsp}
\end{figure}
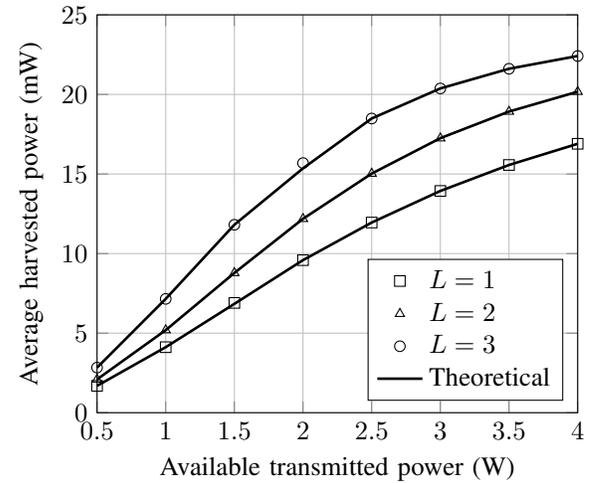

In this subsection, simulations are provided to validate the theoretical results derived in the previous subsection.
In Figs. \ref{fig:ovsd}, \ref{fig:ovsp}, \ref{fig:ahevsd} and \ref{fig:ahevsp}, the performance of the EH system described in section IV is illustrated.
The path loss factor is given by 
$ l = 1 - e^{-\frac{a_ta_r}{(c/f)^2d^2}}$,
where $a_t$ is the aperture of the transmit antenna, $a_r$ the aperture of the receive antenna, $c$ the velocity of light, $f_c$ the operating frequency and $d$ the distance between the transmitter and the receiver. Assuming the receiver as a small sensor, we set $a_t = 0.5$m, $a_r = 0.01$m and $f_c = 2.4$GHz \cite{Tegos2019, Kang2018}.
For the parameters of the nonlinear EH model, we set $A=150$, $B=0.014$ and $P_s=24$mW \cite{Boshkovska2017}. In Figs. \ref{fig:ovsd} and \ref{fig:ovsp}, we normalize the outage threshold $q_t$ with respect to the maximum harvested power when the power harvesting circuit is saturated, termed as $P_s$.
Figs. \ref{fig:ovsd} and \ref{fig:ahevsd} illustrate the performance of the first MISO scenario and the available transmitted power is $2$W and is equally split into the $L$ antennas. Figs. \ref{fig:ovsp} and \ref{fig:ahevsp} illustrate the performance of the second MISO scenario and the distances between the EH source and the three PBs are $12$m, $10$m and $8$m, respectively. The available transmitted power is equally split into the $L$ PBs for fairness. Moreover, it is assumed that $a_i = 3,\ \forall i$ and $b_i = 1, \ \forall i$.

In Figs. \ref{fig:ovsd} and \ref{fig:ovsp}, the outage performance improves as the number of the antennas increases, since if a link undergoes bad channel conditions, outage may not occur due to the other available links. This cannot be the case when only one antenna is used. This is also the reason why the average harvested power increases, as the number of the antennas increases, in Figs. \ref{fig:ahevsd} and \ref{fig:ahevsp}. However, in Fig. \ref{fig:ahevsd} in larger distances the increase of the number of the antennas does not offer an increase in the average harvested power.

\section{Conclusions}

In this work, we utilize the Pearson type III and the log Pearson type III distributions in wireless communications and specifically in WPT, where a frequently-used nonlinear EH model is considered. We extract new closed-form expressions for the statistical properties of a general form of Pearson and log Pearson type III distributions and we utilize them to introduce the logit Pearson type III distribution which is closely related to the specific model and we derive closed form expressions for its CDF, PDF and moments. The distributions of the sum of Pearson type III family RVs and their statistical properties are also investigated. We utilize the derived results to extract closed-form expressions for the CDF, the PDF and the moments of the harvested power for a SISO and a MISO EH system with the considered nonlinear EH model. These statistical properties can provide useful insights such as the probability that outage occurs in the harvested power if we consider a specific threshold and the average harvested power by the source.

\appendices

\section{Proof of Proposition \ref{prop_lpt_mom}} \label{proof_lpt_mom}

The $n$-th moment of the logit Pearson type III distribution when $b>0$ can be obtained by the integral $\int_{\frac{1}{1+e^{-m}}}^{1} z^n f_{Z}(z) dz$ which can be rewritten as
\begin{equation} \label{ltp_mom1}
\mu_Z^n (a,b,m) = \frac{1}{\Gamma(a)} \int_{0}^{\infty} \left( \frac{1}{1+e^{-\frac{z}{b}-m}} \right)^n x^{a-1} e^{-x} dz.
\end{equation}
When $m \geq 0$, it holds that $e^{-\frac{z}{b}-m} \leq 1$ and utilizing the binomial theorem for negative integer exponent, \eqref{ltp_mom1} can be rewritten as
\begin{equation} \label{ltp_mom2}
\begin{split}
\mu_Z^n (a,b,m) & = \frac{1}{\Gamma(a)} \int_{0}^{\infty} \sum_{l=0}^\infty \binom{n+l-1}{l} (-1)^l e^{-m l} \\
	& \times z^{a-1} e^{-\left(1+\frac{l}{b}\right)z} dz.
\end{split}
\end{equation}	
The infinite series in \eqref{ltp_mom2} converges from the utilization of the binomial theorem. When $m \geq 0$, \eqref{ltPT3_moments} is derived by interchanging the order of summations and integrations and using the definition of gamma function and \cite[eq.(3.381.4)]{Gradshteyn2014} .

When $m < 0$, utilizing the binomial theorem for negative integer exponent, the denominator can be written as
\begin{equation} 
\left( 1+e^{-\frac{z}{b}-m} \right)^{-n} \! = \! \begin{cases}
\begin{split}
& \sum_{l=0}^\infty \binom{n+l-1}{l} (-1)^l e^{m (n+l)} e^{\frac{n+l}{b}z}, \\
& \qquad \qquad \qquad x < - m b
\end{split} \\
\begin{split}
& \sum_{l=0}^\infty \binom{n+l-1}{l} (-1)^l e^{-m l} e^{-\frac{l}{b}z}, \\
& \qquad \qquad \qquad x > - m b.
\end{split}
\end{cases}
\end{equation}
In this case, the $n$-th moment can be calculated as 
\begin{equation}
\begin{split}
\mu_Z^n & (a,b,m) = \frac{1}{\Gamma(a)} \int_{0}^{-m b} \sum_{l=0}^\infty \binom{n+l-1}{l} (-1)^l e^{m (n+l)} \\
& \times z^{a-1} e^{-\left(1-\frac{n+l}{b}\right)z} dz 
+ \frac{1}{\Gamma(a)} \int_{-m b}^{\infty} \sum_{l=0}^\infty \binom{n+l-1}{l} \\
& \times (-1)^l e^{-m l} z^{a-1} e^{-\left(1+\frac{l}{b}\right)z} dz.
\end{split}
\end{equation}
Considering the definition of lower and upper incomplete gamma function and \cite[eq.(3.381.1\&3)]{Gradshteyn2014}, \eqref{ltPT3_moments} is derived when $m<0$ which completes the proof.

\section{Proof of Proposition \ref{prop_sum_p}} \label{proof_sum_p}

When $b_i = b$, $\forall i$, the PDF can be derived by the inverse Laplace transform of the product of the moment generation functions of the Pearson type III distribution.	
To derive the PDF of $SX_L$, when $b_i \neq b_j$, $\forall i \neq j$, similar steps as in \cite{Karagiannidis2006} are followed. Firstly, we consider the case of two terms ($L=2$). The PDF of $SX_2 = X_1 + X_2$, considering that if $b>0$, $x\in (m,\infty)$, can be obtained as
\begin{equation} \label{pdf_pr1}
f_{SX_2}(x) \! = \! \int_{m_1}^{z-m_2} \! f_{X_1}(y,a_1,b_1,m_1) f_{X_2}(x-y,a_2,b_2,m_2) dy.
\end{equation}
Using $z=y-m_1$ and \cite[eq.(3.383.1)]{Gradshteyn2014}, \eqref{pdf_pr1} can be rewritten as
\begin{equation} \label{pdf_pr2}
\begin{split}
f_{SX_2}&(x) = \tfrac{b_1^{a_1} b_2^{a_2}}{\Gamma(a_1) \Gamma(a_2)} e^{-b_2(z-m_1-m_2)} (z-m_1-m_2)^{a_1+a_2-1} \\
& \times B(a_2,a_1) {}_1 \! F_1 (a_1 ; a_1 + a_2 ; (b_2-b_1)(z-m_1-m_2)),
\end{split}
\end{equation}
where $B(\cdot,\cdot)$ is the Euler Beta function \cite{Gradshteyn2014} and ${}_1 \! F_1 (\cdot ; \cdot ; \cdot)$ is the confluent hypergeometric function \cite{Gradshteyn2014}. Since $a_i \in \mathbb{Z}$ and $a>0$, using \cite[eq.(07.20.03.0024.01)]{wolfram}, \eqref{pdf_pr2} can be rewritten as 
\begin{equation} \label{pdf_pr3}
\begin{split}
f_{SX_2}&(x) = \frac{(1-a_1-a_2)_{a_1} (b_2-b_1)^{1-a_1-a_2} b_1^{a_1} b_2^{a_2}}{(a_1-1)! (a_1+a_2-1)!}  \\
& \times \left( \sum_{k=0}^{a_2-1} \frac{(1-a_2)_k ((b_2-b_1)(z-m_1-m_2))^k}{k! (2-a_1-a_2)_k} \right. \\
& \times e^{-b_2(z-m_1-m_2)} - e^{-b_1(z-m_1-m_2)} \\
& \left. \times \sum_{k=0}^{a_1-1} \frac{(1-a_1)_k ((b_1-b_2)(z-m_1-m_2))^k}{k! (2-a_1-a_2)_k} \right),
\end{split}
\end{equation}
where $(n)_k$ is the Pochhammer symbol.
After some algebraic manipulations, \eqref{pdf_pr3} can be written as
\begin{equation} \label{pdf_pr4}
f_{SX_2}(x) = \sum_{i=1}^{2} \sum_{k=1}^{a_{i}} f_{X_i}(x,k,b_i,sm_2) 
\Xi_2 \left(i, k, a_1, a_2, b_1, b_2\right),
\end{equation}
where 
\begin{equation} \label{pdf_pr5}
\begin{split}
\Xi_2 &\left(i, k, a_1, a_2, b_1, b_2\right) = (-1)^{sm_2 - m_i} \frac{b_1^{a_1} b_2^{a_2}}{b_k} \\
& \times \frac{(a_1+a_2-k-1)! \left(b_i-b_{1+U(1-i)}\right)^{k-a_1-a_2}}{\left(a_{1+U(1-i)}-1\right)! (a_i-k)!} .
\end{split}
\end{equation}
It should be highlighted that if $b<0$, $x\in (-\infty,m)$ and the PDF of $SX_2$ can be obtained as
\begin{equation} \label{pdf_pr1.2}
f_{SX_2}(x) \! = \! \int_{z-m_2}^{m_1} \! f_{X_1}(y,a_1,b_1,m_1) f_{X_2}(x-y,a_2,b_2,m_2) dy,
\end{equation}
which results in \eqref{pdf_pr4} considering that $|b|$ appears in \eqref{PT3_PDF}.

When $L=3$, the PDF of $SX_3 = SX_2 + X_3$ can be obtained as
\begin{equation} \label{pdf_pr6}
f_{SX_3}(x) \! = \! \int_{m_1+m_2}^{z-m_3} \! f_{SX_2}(y) f_{X_3}(x-y,a_3,b_3,m_3) dy.
\end{equation}
Following similar steps and after some complicated algebraic manipulations, \eqref{pdf_pr6} can written as
\begin{equation} \label{pdf_pr7}
\begin{split}
f_{SX_3}(x) & = \sum_{i=1}^{3} \sum_{k=1}^{a_{i}} f_{X_i}(x,k,b_i,sm_3) \\
& \times \Xi_3 \left(i, k, a_1, a_2, a_3, b_1, b_2, b_3\right),
\end{split}
\end{equation}
where 
\begin{equation} \label{pdf_pr8}
\begin{split}
\Xi_3 &\left(i, k, a_1, a_2, a_3, b_1, b_2, b_3, j_1\right) = (-1)^{sm_3 - m_i} \frac{b_1^{a_1} b_2^{a_2} b_3^{a_3}}{b_k} \\
& \times \tfrac{\left(a_{i}+a_{1+U(1-i)}-j_{1}-1\right)! \left(b_i-b_{1+U(1-i)}\right)^{j_{1}-a_{i}-a_{1+U(1-i)}}}{\left(a_{1+U(1-i)}-1\right)! \left(a_{i}-j_{1}\right)!} \\
& \times \tfrac{\left( j_{1}+a_{2+U(2-i)}-k-1 \right)! \left(b_{i} - b_{2+U(2-i)}\right)^{k-j_{1}-a_{2+U(2-i)}}}{ \left(a_{2+U(2-i)}-1\right)! \left(j_{1}-k\right)!} .
\end{split}
\end{equation}
When $b_i<0$, $\forall i$, the PDF of $SX_3$ can be obtained as
\begin{equation} \label{pdf_pr6.2}
f_{SX_3}(x) \! = \! \int_{z-m_3}^{m_1+m_2} \! f_{SX_2}(y) f_{X_3}(x-y,a_3,b_3,m_3) dy,
\end{equation}
which results in \eqref{pdf_pr7}.

Following similar steps for $L$ terms, \eqref{pdfSX} is derived.

\section{Proof of Theorem \ref{th_EH_moments}} \label{proof_EH_moments}

Setting in $\int_0^{P_s} q^n f_{Q^S}(q) dq$ 
\begin{equation}
x = - \hat{b} \left( \ln \left( \frac{c\left(1+e^{AB}\right)}{q + c} - 1 \right) - A B \right),
\end{equation}
the $n$-th moment is calculated as
\begin{equation} \label{EH_mom_pr1}
\mu_{Q^S}^n = \frac{c^n}{\Gamma (a)} \int_0^{\infty} e^{-x} x^{a-1} \left(\frac{e^{-A B}+1}{e^{- x / \hat{b}}+e^{-A B}}-1\right)^n dx.
\end{equation}
Using the binomial theorem, \eqref{EH_mom_pr1} can be rewritten as
\begin{equation} \label{EH_mom_pr2}
\begin{split}
\mu_{Q^S}^n & = \frac{c^n}{\Gamma(a)} \sum _{l_1=0}^n \binom{n}{l_1} (-1)^{n-l_1} \left(e^{-A B}+1\right)^{l_1} \\
& \times \int_0^{\infty} \frac{x^{a-1} e^{-x}}{\left(e^{- x / \hat{b}}+e^{-A B}\right)^{l_1}} dx.
\end{split}
\end{equation}
Utilizing the binomial theorem for negative integer exponent, the denominator can be written as
\begin{equation} \label{neg_bin}
\left(e^{- x / \hat{b}}+e^{-A B}\right)^{-l_1} \! = \! \begin{cases}
\begin{split}
& \sum _{l_2=0}^\infty \binom{l_1+l_2-1}{l_2} \left(e^{- x / \hat{b}}\right)^{-l_1-l_2} \\
& \quad \times (-1)^{l_2} \left(e^{-A B}\right)^{l_2}, \ x < A B \hat{b} 
\end{split} \\
\begin{split}
& \sum _{l_2=0}^\infty \binom{l_1+l_2-1}{l_2} \left(e^{-A B}\right)^{-l_1-l_2} \\
& \quad \times (-1)^{l_2} \left(e^{- x / \hat{b}}\right)^{l_2} , \ x > A B \hat{b}.
\end{split}
\end{cases}
\end{equation}
In this case, the infinite series always converge.
Using \eqref{neg_bin}, \eqref{EH_mom_pr2} can be rewritten as
\begin{equation} \label{integral_f}
\begin{split}
\mu_{Q^S}^n & = \frac{c^n}{\Gamma(a)} \sum_{l_1=0}^n \sum_{l_2=0}^\infty \binom{n}{l_1} \binom{l_1+l_2-1}{l_2} (-1)^{n-l_1+l_2} \\
& \times \left( e^{-A B l_2} \int_0^{A B \hat{b}} x^{a-1} e^{-( 1 - (l_1+l_2)/\hat{b} )x} dx \right. \\
& \left. + e^{A B (l_1+l_2)} \int_{A B \hat{b}}^\infty x^{a-1} e^{-( 1 + l_2/\hat{b} )x} dx \right) \left(e^{-A B}+1\right)^{l_1}.
\end{split}
\end{equation}
Considering the definition of lower and upper incomplete gamma function and \cite[eq.(3.381.1\&3)]{Gradshteyn2014}, respectively, \eqref{EH_moments} is derived.

\bibliographystyle{IEEEtran}
\bibliography{Bibliography}

\begin{thebibliography}{10}
\providecommand{\url}[1]{#1}
\csname url@samestyle\endcsname
\providecommand{\newblock}{\relax}
\providecommand{\bibinfo}[2]{#2}
\providecommand{\BIBentrySTDinterwordspacing}{\spaceskip=0pt\relax}
\providecommand{\BIBentryALTinterwordstretchfactor}{4}
\providecommand{\BIBentryALTinterwordspacing}{\spaceskip=\fontdimen2\font plus
\BIBentryALTinterwordstretchfactor\fontdimen3\font minus
  \fontdimen4\font\relax}
\providecommand{\BIBforeignlanguage}[2]{{%
\expandafter\ifx\csname l@#1\endcsname\relax
\typeout{** WARNING: IEEEtran.bst: No hyphenation pattern has been}%
\typeout{** loaded for the language `#1'. Using the pattern for}%
\typeout{** the default language instead.}%
\else
\language=\csname l@#1\endcsname
\fi
#2}}
\providecommand{\BIBdecl}{\relax}
\BIBdecl

\bibitem{Johnson1995}
N.~L. Johnson, S.~Kotz, and N.~Balakrishnan, \emph{Continuous univariate
  distributions}.\hskip 1em plus 0.5em minus 0.4em\relax John Wiley \& Sons,
  Ltd, 1995.

\bibitem{Singh1998_p}
V.~P. Singh, ``{Pearson Type III Distribution},'' in \emph{Entropy-Based
  Parameter Estimation in Hydrology}.\hskip 1em plus 0.5em minus 0.4em\relax
  Springer, Jan. 1998.

\bibitem{Singh1998_lp}
------, ``{Log-Pearson Type III Distribution},'' in \emph{Entropy-Based
  Parameter Estimation in Hydrology}.\hskip 1em plus 0.5em minus 0.4em\relax
  Springer, Jan. 1998.

\bibitem{Bobee1975}
B.~Bobee, ``{The log Pearson type 3 distribution and its application in
  hydrology},'' \emph{Water resources research}, vol.~11, no.~5, pp. 681--689,
  1975.

\bibitem{Phien1984}
H.~N. Phien and T.~J. Ajirajah, ``{Applications of the log Pearson type-3
  distribution in hydrology},'' \emph{Journal of Hydrology}, vol.~73, no.~3,
  pp. 359--372, 1984.

\bibitem{Gupta1994}
I.~Gupta and V.~Deshpande, ``{Application of log-Pearson type-Ill distribution
  for evaluating design earthquake magnitudes},'' \emph{Journal of the
  Institution of Engineers (India), Civil Engineering Division}, vol.~75, pp.
  129--134, 1994.

\bibitem{Shi2017}
Z.~{Shi}, S.~{Ma}, G.~{Yang}, K.~{Tam}, and M.~{Xia}, ``{Asymptotic Outage
  Analysis of HARQ-IR Over Time-Correlated Nakagami- $m$ Fading Channels},''
  \emph{IEEE Trans. Wireless Commun.}, vol.~16, no.~9, pp. 6119--6134, Jun.
  2017.

\bibitem{Kim2003}
S.~{Kim}, J.~Y. {Lee}, and D.~K. {Sung}, ``{A shifted gamma distribution model
  for long-range dependent Internet traffic},'' \emph{IEEE Commun. Lett.},
  vol.~7, no.~3, pp. 124--126, Mar. 2003.

\bibitem{diamantoulakis2018resource}
P.~D. Diamantoulakis, ``Resource allocation in wireless networks with energy
  constraints,'' Ph.D. dissertation, Aristotle Univeristy of Thessaloniki,
  Thessaloniki, Greece, 2017.

\bibitem{Grover2010}
P.~Grover and A.~Sahai, ``{Shannon Meets Tesla: Wireless Information and Power
  Transfer},'' in \emph{Proc. IEEE Int. Symp. Information Theory}, Austin, TX,
  USA, Jun. 2010, pp. 2363--2367.

\bibitem{Krikidis2012}
I.~{Krikidis}, S.~{Timotheou}, and S.~{Sasaki}, ``{RF Energy Transfer for
  Cooperative Networks: Data Relaying or Energy Harvesting?}'' \emph{IEEE
  Commun. Lett.}, vol.~16, no.~11, pp. 1772--1775, Sep. 2012.

\bibitem{Krikidis2014}
I.~Krikidis, S.~Timotheou, S.~Nikolaou, G.~Zheng, D.~W.~K. Ng, and R.~Schober,
  ``Simultaneous wireless information and power transfer in modern
  communication systems,'' \emph{IEEE Commun. Mag.}, vol.~52, no.~11, pp.
  104--110, Nov. 2014.

\bibitem{Krikidis2020}
I.~{Krikidis}, ``{Wireless Power Transfer Under Kullback-Leibler Distribution
  Uncertainty: A Mathematical Framework},'' \emph{IEEE Wireless Commun. Lett.},
  vol.~9, no.~9, pp. 1591--1595, Sept. 2020.

\bibitem{Tegos2020}
S.~A. {Tegos}, P.~D. {Diamantoulakis}, A.~S. {Lioumpas}, P.~G. {Sarigiannidis},
  and G.~K. {Karagiannidis}, ``{Slotted ALOHA With NOMA for the Next Generation
  IoT},'' \emph{IEEE Trans. Commun.}, vol.~68, no.~10, pp. 6289--6301, Oct.
  2020.

\bibitem{Zhou2014}
X.~Zhou, R.~Zhang, and C.~K. Ho, ``{Wireless Information and Power Transfer in
  Multiuser OFDM Systems},'' \emph{IEEE Trans. Wireless Commun.}, vol.~13,
  no.~4, pp. 2282--2294, Apr. 2014.

\bibitem{Diamantoulakis2016}
P.~D. Diamantoulakis, K.~N. Pappi, Z.~Ding, and G.~K. Karagiannidis,
  ``{Wireless-Powered Communications With Non-Orthogonal Multiple Access},''
  \emph{IEEE Trans. Wireless Commun.}, vol.~15, no.~12, pp. 8422--8436, Dec.
  2016.

\bibitem{Zhang2013}
R.~Zhang and C.~K. Ho, ``{MIMO Broadcasting for Simultaneous Wireless
  Information and Power Transfer},'' \emph{IEEE Trans. Wireless Commun.},
  vol.~12, no.~5, pp. 1989--2001, May 2013.

\bibitem{Chen2017}
Y.~Chen, N.~Zhao, and M.-S. Alouini, ``{Wireless Energy Harvesting Using
  Signals From Multiple Fading Channels},'' \emph{IEEE Trans. Commun.},
  vol.~65, pp. 5027--5039, Nov. 2017.

\bibitem{Clerckx2018}
B.~Clerckx, ``{Wireless Information and Power Transfer: Nonlinearity, Waveform
  Design, and Rate-Energy Tradeoff},'' \emph{IEEE Trans. Signal Process.},
  vol.~66, pp. 847--862, Feb. 2018.

\bibitem{Boshkovska2015}
E.~{Boshkovska}, D.~W.~K. {Ng}, N.~{Zlatanov}, and R.~{Schober}, ``{Practical
  Non-Linear Energy Harvesting Model and Resource Allocation for SWIPT
  Systems},'' \emph{IEEE Commun. Lett.}, vol.~19, no.~12, pp. 2082--2085, Dec.
  2015.

\bibitem{Kang2018}
J.~Kang, I.~Kim, and D.~I. Kim, ``{Wireless Information and Power Transfer:
  Rate-Energy Tradeoff for Nonlinear Energy Harvesting},'' \emph{IEEE Trans.
  Wireless Commun.}, vol.~17, no.~3, pp. 1966--1981, Mar. 2018.

\bibitem{Clerckx2019}
B.~Clerckx, R.~Zhang, R.~Schober, D.~W.~K. Ng, D.~I. Kim, and H.~V. Poor,
  ``{Fundamentals of Wireless Information and Power Transfer: From RF Energy
  Harvester Models to Signal and System Designs},'' \emph{IEEE J. Sel. Areas
  Commun.}, vol.~37, no.~1, pp. 4--33, Jan. 2019.

\bibitem{Tegos2019}
S.~A. {Tegos}, P.~D. {Diamantoulakis}, K.~N. {Pappi}, P.~C. {Sofotasios},
  S.~{Muhaidat}, and G.~K. {Karagiannidis}, ``{Toward Efficient Integration of
  Information and Energy Reception},'' \emph{IEEE Trans. Commun.}, vol.~67,
  no.~9, pp. 6572--6585, Sep. 2019.

\bibitem{Jiang2020}
R.~{Jiang}, K.~{Xiong}, P.~{Fan}, Z.~{Zhong}, and K.~B. {Letaief},
  ``{Information-Energy Region for SWIPT Networks in Mobility Scenarios},''
  \emph{IEEE Trans. Veh. Technol.}, vol.~69, no.~7, pp. 7264--7280, 2020.

\bibitem{Johnson1949}
\BIBentryALTinterwordspacing
N.~L. Johnson, ``{Systems Of Frequency Curves Generated by Methods of
  Translation},'' \emph{Biometrika}, vol.~36, no. 1-2, pp. 149--176, 06 1949.
  [Online]. Available: \url{https://doi.org/10.1093/biomet/36.1-2.149}
\BIBentrySTDinterwordspacing

\bibitem{Frederic2008}
P.~Frederic and F.~Lad, ``Two moments of the logitnormal distribution,''
  \emph{{Communications in Statistics—Simulation and Computation}}, vol.~37,
  no.~7, pp. 1263--1269, 2008.

\bibitem{Karagiannidis2004}
G.~K. {Karagiannidis}, ``{Moments-based approach to the performance analysis of
  equal gain diversity in Nakagami-m fading},'' \emph{IEEE Trans. Commun.},
  vol.~52, no.~5, pp. 685--690, May 2004.

\bibitem{Coelho1998}
C.~A. Coelho, ``{The generalized integer Gamma distribution—a basis for
  distributions in multivariate statistics},'' \emph{Journal of Multivariate
  Analysis}, vol.~64, no.~1, pp. 86--102, 1998.

\bibitem{Karagiannidis2006}
G.~K. {Karagiannidis}, N.~C. {Sagias}, and T.~A. {Tsiftsis}, ``{Closed-form
  statistics for the sum of squared Nakagami-m variates and its
  applications},'' \emph{IEEE Trans. Commun.}, vol.~54, no.~8, pp. 1353--1359,
  Aug. 2006.

\bibitem{Gradshteyn2014}
I.~S. Gradshteyn and I.~M. Ryzhik, \emph{Table of integrals, series, and
  products}.\hskip 1em plus 0.5em minus 0.4em\relax Academic press, 2014.

\bibitem{Boshkovska2017}
E.~{Boshkovska}, N.~{Zlatanov}, L.~{Dai}, D.~W.~K. {Ng}, and R.~{Schober},
  ``{Secure SWIPT Networks Based on a Non-Linear Energy Harvesting Model},'' in
  \emph{2017 IEEE Wireless Communications and Networking Conference Workshops
  (WCNCW)}, San Francisco, CA, USA, Mar. 2017, pp. 1--6.

\bibitem{wolfram}
{The Wolfram Functions Site [Online]}, ``{Available:
  \url{http://functions.wolfram.com}}.''

\end{thebibliography}


\end{document}